\documentclass[showpacs,aps,epsfig,nofootinbib]{revtex4}
\usepackage{mathrsfs}

\usepackage{graphicx}
\newcommand{\ba}{\begin{eqnarray}} \newcommand{\ea}{\end{eqnarray}}
\usepackage{amsfonts}
\usepackage{epstopdf}
\usepackage{latexsym}
\usepackage{amssymb}
\usepackage{amssymb}

\usepackage[center]{subfigure}

\begin{document}

 \newcommand{\bq}{\begin{equation}}
 \newcommand{\eq}{\end{equation}}
 \newcommand{\bqn}{\begin{eqnarray}}
 \newcommand{\eqn}{\end{eqnarray}}
 \newcommand{\nb}{\nonumber}
 \newcommand{\lb}{\label}
\newcommand{\PRL}{Phys. Rev. Lett.}
\newcommand{\PL}{Phys. Lett.}
\newcommand{\PR}{Phys. Rev.}
\newcommand{\CQG}{Class. Quantum Grav.}

\title{Nonlinear Electromagnetic Quasinormal Modes and Hawking Radiation of A Regular Black Hole with Magnetic Charge}

\author{Jin Li}
\email{cqstarv@hotmail.com}
\affiliation {Department of Physics, Chongqing University,
Chongqing 400030, China}
\
\author{Kai Lin}
\email{lk314159@hotmail.com}
\affiliation {Instituto de F\'isica, Universidade de S\~ao Paulo, CP 66318, 05315-970, S\~ao Paulo, Brazil}
\author{Nan Yang}
\email{cqunanyang@hotmail.com}
\affiliation {Department of Physics, Huazhong University of Science and Technology, Wuhan 430074, China}
\date{\today}

\begin{abstract}
Based on a regular exact black hole (BH) from nonlinear electrodynamics
(NED) coupled to General Relativity, we investigate its stability of such BH through the Quasinormal Modes (QNMs) of
electromagnetic (EM) field perturbation and its thermodynamics through Hawking radiation. In perturbation theory, we can deduce the effective potential from nonlinear EM field. The comparison of potential function between regular and RN BHs could predict their similar QNMs. The QNMs
frequencies tell us the effect of magnetic charge $q$, overtone $n$,
angular momentum number $l$ on the dynamic evolution of NLED EM
field. Furthermore we also discuss the cases near extreme condition of such magnetically charged
regular BH. The corresponding QNMs spectrum illuminates some special
properties in the near-extreme cases. For the thermodynamics, we employ Hamilton-Jacobi method to calculate the near-horizon Hawking temperature of the regular BH and reveal the relationship between classical parameters of black hole and its quantum effect.
\\
\\
\textbf{Keywords}:regular black hole; nonlinear electrodynamics; Quasinormal Modes

\end{abstract}

\pacs{04.70.Bw; 04.62.+v}

\maketitle
\section{introduction}

In the framework of black hole, the research of singularity-free black holes has attracted a considerable attention during the last decades. Fortunately physicists have found a new type of black-hole solutions without singularity in General Relativity~\cite{addnewFRK} and
more general gravity theories to
construct solutions without a singularity, such as string theory
and exact conformal field theory~\cite{intref1,intref2}. Especially when gravitation coupling to a suitable nonlinear electrodynamics (NLED) field, some regular metrics would be obtained~\cite{addadd1}-\cite{intref5}. Ayon-Beato,
Garcia and Bronnikov successively found some static, spherically symmetric non-singular solutions~\cite{mainref}-\cite{intref5}. It is also remarkable that all the
NLED satisfy the zeroth and first laws of BH
mechanics~\cite{intref6}. In the theory of NLED coupling to Einstein
equations, it is important to seek a suitable gauge-invariant
Lagrangians ${\cal L}(F)$ ($F=F_{\mu\nu}F^{\mu\nu}$ is the electromagnetic
tensor), and its energy-momentum tensor (EMT)
\begin{equation}
T_{\nu\mu}=\frac{1}{4\pi}\left[\frac{d{\cal L}(F)}{dF}F_{\rho\mu}F^{\rho}_{\nu}-\frac{1}{4}g_{\mu\nu}{\cal L}(F)\right],
\end{equation}
satisfying the symmetry $T^{0}_{0}=T^{1}_{1}$~\cite{addref1,mainref}. In this
paper, we consider the stability of a specific regular BH solution, supposing
\begin{equation}
{\cal L}(F)=F\cosh^{-2}\left(a\left|\frac{F}{2}\right|^{1/4}\right),\label{L}
\end{equation}
in NLED thoery. where $a=|q|^{\frac{3}{2}}/2M$ ($M$ is the mass of BH). Without loss the generality, if there is no specific notation we set $M=1$ in this paper. Then the static, spherically symmetric metric is described as~\cite{mainref,intref4}.
\begin{equation}
\lb{1}
ds^{2}=-f(r)dt^{2}+f(r)^{-1}dr^{2}+r^{2}d\theta^{2}+r^{2}\sin^{2}\theta
d\phi^{2},\label{eq:metric}
\end{equation}
where
\begin{equation}
f(r)=1-\frac{2M}{r}\left[1-\tanh\left(\frac{q^{2}}{2Mr}\right)\right],
\end{equation}
where $q$ should obey Theorem 1 discussed in~\cite{mainref}
(i.e., $q=q_{m}\neq0,q_{e}=0$, where $q_{m}$ is magnetic charge). About a decade ago, Mosquera Cuesta and Salim presented with NLED the gravitational redshift of super-strongly magnetized compact objects, such as pulsars and particular neutron stars would rely definitely on the magnetic field permeating the magnetized objects\cite{ref1-1,ref1-2}. That opposes to general relativity, where the gravitational redshift and background magnetic field are unrelated. Meanwhile they found gravitational redshift is related to the mass-radius ratio of the object, so that this NLED effect would also impact on the metric of black hole\cite{ref1-1,ref1-2}. The discussion of effective metric of our regular spacetime indicates the original metric is indeed revised by NLED moderately (see Appendix $\rm{A}$). Since the discrepancy mainly concentrates on the area around $r=0$, we still consider the EM perturbation in the original spacetime in the following context.

A black hole can be described completely under perturbation, so that it is necessary to analyze the perturbations once you want to know something about its stability~\cite{intref7}. After perturbation,
BH will experience three stages (Initial oscillation - QNMs ringing - Ringdown), which can give us a glimpse into
the interior region of black holes. The second stage named
``quasinormal modes (QNMs) ringing'' contributes to
gravitational wave (GW) detection. From a theoretical point of view,
perturbations of a black hole space-time can be performed in two
ways: by adding fields to the black hole space-time or by perturbing
the black hole metric (the background) itself. The former way
refers to physical particle fields such as scalar, Dirac
and electromagnetic(EM) field; The latter one is the
gravitational perturbation resulting in GWs. Many researches into regular BH QNMs have made great progress, focusing
on scalar field \cite{RegularBH,ourIJTP,citeour1,citeour2} and Dirac perturbations
\cite{ourpaper}. Usually the singularity of a regular black hole can be vanished under the condition when gravitation coupling to a suitable nonlinear electrodynamics (NLED) field, therefore we consider the QNMs of EM perturbations in specific regular space-times will be more meaningful.

The beginning of a QNMs calculation is to reduce the perturbation
equations (which vary with the spin of perturbation fields) into the
two-dimensional wavelike form with decoupled angular variables. Once
the variables are decoupled, the equation for radial and
time variables usually has the Schr\"odinger-like form in
stationary background. Then the corresponding potential function
$V(r)$ can be determined, which is the key to numerical computation of QNMs frequency (QNFs). The numerical methods for calculation of QNFs
have been developed for several years, and now mainly consist of time
domain method~\cite{QNMmethod1, QNMmethod2},expansion
method~\cite{eikonal1,eikonal2},direct integration in the frequency
domain~\cite{QNMmethod3},
WKB method~\cite{QNMmethod4,QNMmethod5,QNMmethod6,QNMmethod7},finite
differential method~\cite{Finit1,Finit2} etc. Since the WKB scheme
has been shown to be more accurate for both the real and imaginary
parts of the dominant QNMs with $n\leq l$~\cite{Diracfield}, we
apply WKB method to the QNFs calculation and compare the results with the ones
from expansion method.

As important as the stability of BH, the thermodynamics of black holes is thought to be the connection between black hole physics and quantum theory. Hawking radiation can be seen as quantum tunneling around the horizon, and the Hawking temperature of black hole can be deduced through tunneling rate~\cite{addref2,addref3,tun1,tun2,tun3,tun5}. In order to find the effect of their classical parameters on Hawking radiation, we also study this topic for the magnetically charged regular BH in this paper.

The paper is organized as follows. In section~\ref{sec:regNED}, we
describe the nonlinear electromagnetic field equations in
regular spacetimes and determine the shape of potential. Since the
regular space-time asymptotically behaves as the Reissner Nordstr\"{o}m(RN) BH, we
compare them. In
section~\ref{sec:regQNM} we use $6^{\rm th}$-order WKB method and expansion
method to compute the QNFs of the regular solution and RN BH.
Furthermore we apply finite differential method to display intuitive images of the QNM perturbation. In
section~\ref{sec:extremecondition}, we investigate the strong charged cases for the spherically symmetric regular black hole and
evaluate the QNMs through $3^{\rm rd}$-order WKB method, since the
6$^{th}$ order WKB method may break down if the potential is
complicated~\cite{eikonal1,eikonal2}. In order to further understand the magnetically charged regular BH, Section~\ref{sec:Hawking} shows the Hawking radiation of such BH. Conclusions and future work are presented
in section~\ref{sec:conclusions}.

\section{The nonlinear electromagnetic field perturbation to the regular BH}\label{sec:regNED}
   The action proposed in Einstein-dual nonlinear electromagnetic theory is~\cite{action}
\begin{equation}
\lb{2} S=\frac{1}{16\pi}\int d^{4}x\sqrt{-g}\left[R-{\cal L}(F)\right],
\end{equation}
where $R$ is the scalar curvature, $F=F_{\mu\nu}F^{\mu\nu}, F_{\mu\nu}=\partial_{\mu}A_{\nu}-\partial_{\nu}A_{\mu}$ is the electromagnetic field. ${\cal L}(F)$ is the Lagrangian function of this theory, and returns to the linear case (i.e.,Maxwell field) at small $F$: ${\cal L}(F)\simeq F$ as $F\rightarrow0$. The tensor $F_{\mu\nu}$ obeys the equation~\cite{mainref}
\begin{equation}
\lb{3} \nabla_{\mu}\left({\cal L}_{F}F^{\mu\nu}\right)=0,
\end{equation}
with ${\cal L}_{F}=dL/dF$. We consider the background to be a spherically
symmetric regular space-time involving only a radial magnetic field
(i.e., $\bar{A}_{3}=-q_{m}\cos{\theta}$ and other components of
$\bar{A}_{\mu}$ equal to zero)~\cite{A3}. The zero-order solution of Eq.(\ref{3}) is the solution of background EM field in the regular static spacetime. Then we add the perturbation,
$\delta A_{3}$, to the background EM field, yielding
\begin{equation}
\lb{Abar} A_{3}=\bar{A}_{3}+\delta A_3,
\end{equation}
where the NLED electromagnetic
perturbation, $\delta A_3$, has a small value , which can be expressed as
$\psi(t,r)P_{l}$ ($P_{l}=P_{l}(\text{cos}\theta)$ is Legendre function) after separation of variables. According to
$F_{\mu\nu}=\partial_{\mu}A_{\nu}-\partial_{\nu}A_{\mu}$,
$F^{\mu\nu}=F_{\alpha\beta}g^{\mu\alpha}g^{\beta\nu}$ and neglecting the second and higher orders perturbed terms, we can
get the non-zero perturbed $F^{\mu\nu}$ and $F$ to be
\bqn\label{Fp}
F^{03}=-F^{30}=-\frac{\text{csc}^{2}\theta P_{l}}{r^{2}f(r)}\frac{\partial\psi(t,r)}{\partial t},\nb\\
F^{13}=-F^{31}=\frac{\text{csc}^{2}\theta f(r)P_{l}}{r^{2}}\frac{\partial\psi(t,r)}{\partial r},\\
F^{23}=-F^{32}=\frac{\text{csc}^{2}\theta}{r^{4}}(q\text{sin}\theta+\psi(t,r)\frac{dP_{l}}{d\theta}).\nb
\eqn
\begin{equation}
F=\frac{2q^{2}}{r^{4}}+\frac{4q\text{csc}\theta\psi(t,r)}{r^{4}}\frac{dP_{l}}{d\theta}.\label{Fwith perturbation}
\end{equation}
Substituting Eq.(\ref{Fwith perturbation}) into Eq.(\ref{L}) and supposing $\psi(t,r)=\phi(r)\exp(-i\omega t)$ yield
${\cal L}_{F}$ as
\begin{equation}
{\cal L}_{F}=\bar{\cal L}_{F}+\delta{\cal L}_{F},
\end{equation}
where $\bar{\cal L}_{F}$ is the ${\cal L}_{F}$ of the background EM field and $\delta{\cal L}_{F}$ is the perturbation term of $\cal L_{F}$. They are
\ba
\bar{\cal L}_{F}=-\frac{1}{2}\text{sech}^{2}(\frac{a\sqrt q}{r})\left(-2+\frac{a\sqrt q}{r}\text{tanh}(\frac{a\sqrt q}{r})\right),
\ea
\ba
\delta{\cal L}_{F}=\frac{1}{8q}\left[\frac{a\sqrt q}{r}\text{csc}\theta\text{sech}^{4}(\frac{a\sqrt q}{r})\left(-4\frac{a\sqrt q}{r}+2\frac{a\sqrt q}{r}\text{cosh}(\frac{a\sqrt q}{r})-5\text{sinh}(2\frac{a\sqrt q}{r})\right)\frac{dP_{l}}{d\theta}\phi(r)\exp(-i\omega t)\right].
\ea
Therefore, substituting above quantities into Eq.(\ref{3}) and defining $\phi(r)=B(r)R(r)$ (where $B(r)$ satisfying $B'(r)/B(r)=-\bar{\cal L}'_{F}/2\bar{\cal L}_{F}$), we obtain the
main equation with the first order in the perturbation
\ba\label{mainEq}
\left[\frac{d^{2}}{dr^{2}_{\ast}}+\omega^{2}-V_{\rm{regular}}(r)\right]R(r)=0,
\ea
where``~$'$~" represents $d/dr$, $d/dr_{*}=f(r)d/dr$ and
\ba
V_{\rm{regular}}(r)&=&\frac{l(l+1)f(r)}{r^{2}}-\frac{f^{3}\bar{\cal L}'_{F}\left(\bar{\cal L}_{F}f'+f\bar{\cal L}'_{F}\right)\left(-3\bar{\cal L}'^{2}_{F}+2\bar{\cal L}_{F}\bar{\cal L}''_{F}\right)}{8\bar{\cal L}^{4}_{F}}
 \nonumber\\ &&
 -\frac{f(r)}{4r^{4}\bar{\cal L}_{F}}\left\{a^{2}ql(l+1)\text{sech}^{2}\left(\frac{a\sqrt{q}}{r}\right)\left[-2+3\text{sech}^{2}\left(\frac{a\sqrt{q}}{r}\right)
+5\frac{r}{a\sqrt{q}}\tanh\left(\frac{a\sqrt{q}}{r}\right)\right]\right\}.
 \ea

The RN BH is a result from the linear electromagnetic field, in which $q$ represents the electric charge $q_{e}$. So it is meaningful to compare the effective potential function $V(r)$ with the RN solution. We use
the same principle under the following replacement $A_0=q/r$;
$A_{3}=\delta A_3$; $F=-2q^{2}/r^{4}$;${\cal L}(F)=F$,
yielding
 \ba &&
V_{RN}(r)=\frac{l(l+1)f_{RN}(r)}{r^{2}}-\frac{f_{RN}^{3}\bar{\cal L}'_{F}(\bar{\cal L}_{F}f_{RN}'+f_{RN}\bar{\cal L}'_{F})(-3\bar{\cal L}'^{2}_{F}+2\bar{\cal L}_{F}\bar{\cal L}''_{F})}{8\bar{\cal L}^{4}_{F}}.\label{VRN}
 \ea
Note: (1) Due to ${\cal L}(F)=F$, $\bar{{\cal L}}_{F}=1$, the
above equation can be simplified as
\begin{equation}
V_{RN}(r)=\frac{l(l+1)f_{RN}(r)}{r^{2}}.
\end{equation}
(2) Actually as the result of linear electromagnetic field, there is a category of charged black holes named Kerr-Newman-Kusuya black hole\cite{NK0,NK1,NK2}, which represents a number of static or stationary rotating black holes with electric and magnetic charges. But using an equivalent charge $q_{h}$ discussed in Appendix $\rm{C}$ instead of $q_{e}$, the spacetime of Kerr-Newman-Kusuya black hole without rotating has the same expression as ordinary RN solution. So we choose RN as a representative to be discussed in this paper.
\begin{figure}
\centering \subfigure{
\begin{minipage}[t]{0.3\textwidth}
\includegraphics[width=1\textwidth]{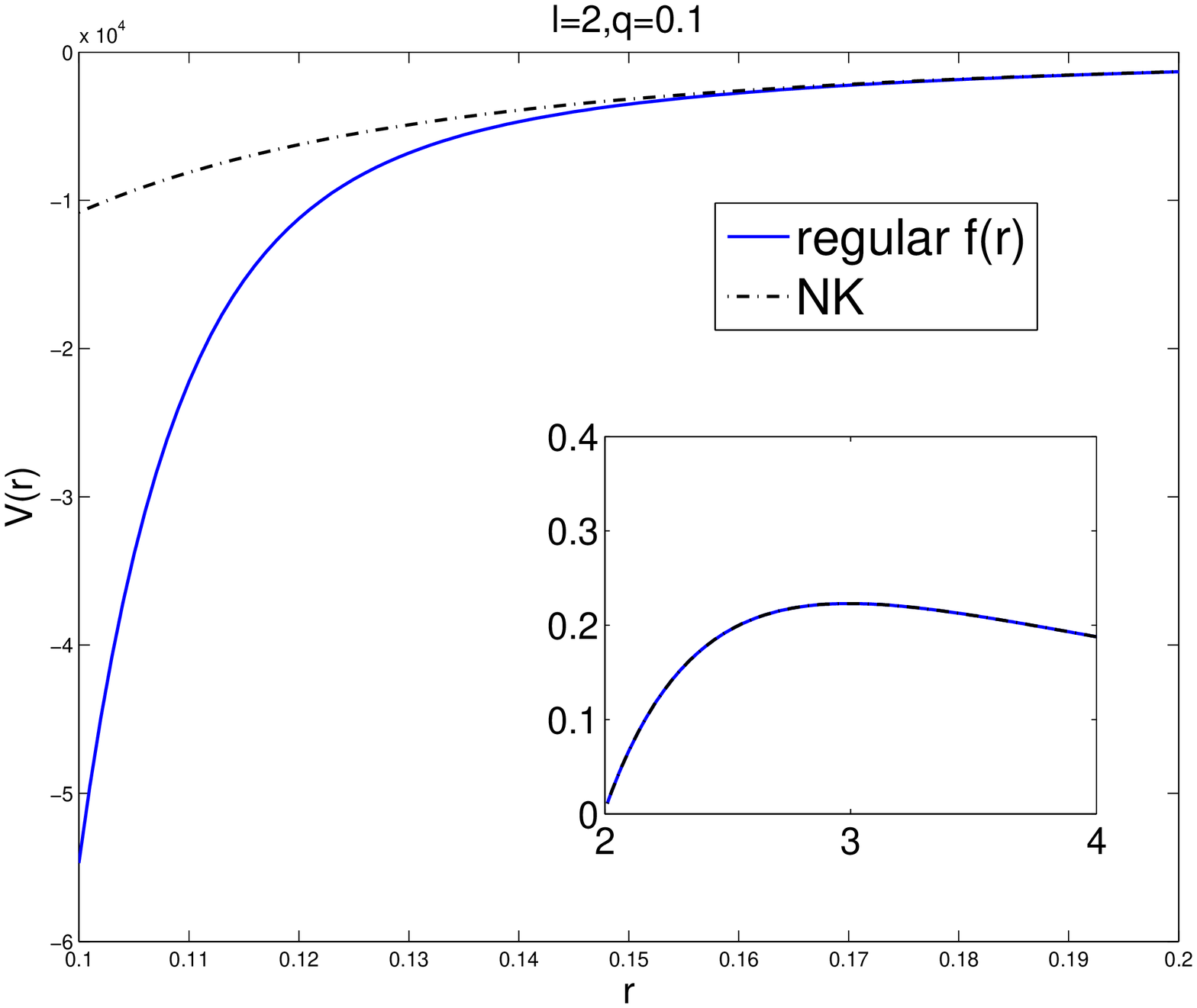}
\end{minipage}
} \subfigure{
\begin{minipage}[t]{0.3\textwidth}
\includegraphics[width=1\textwidth]{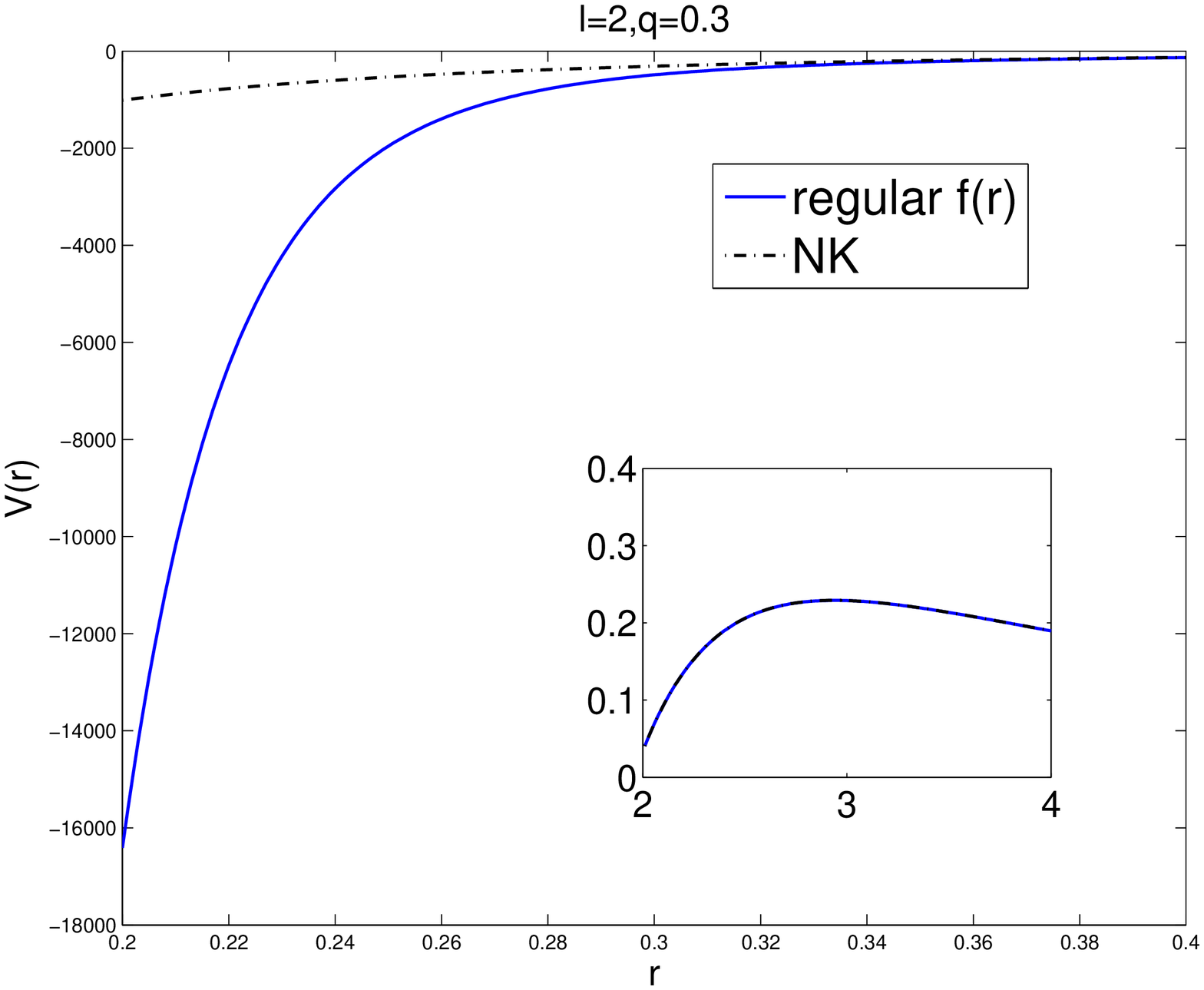}
\end{minipage}
} \subfigure{
\begin{minipage}[t]{0.3\textwidth}
\includegraphics[width=1\textwidth]{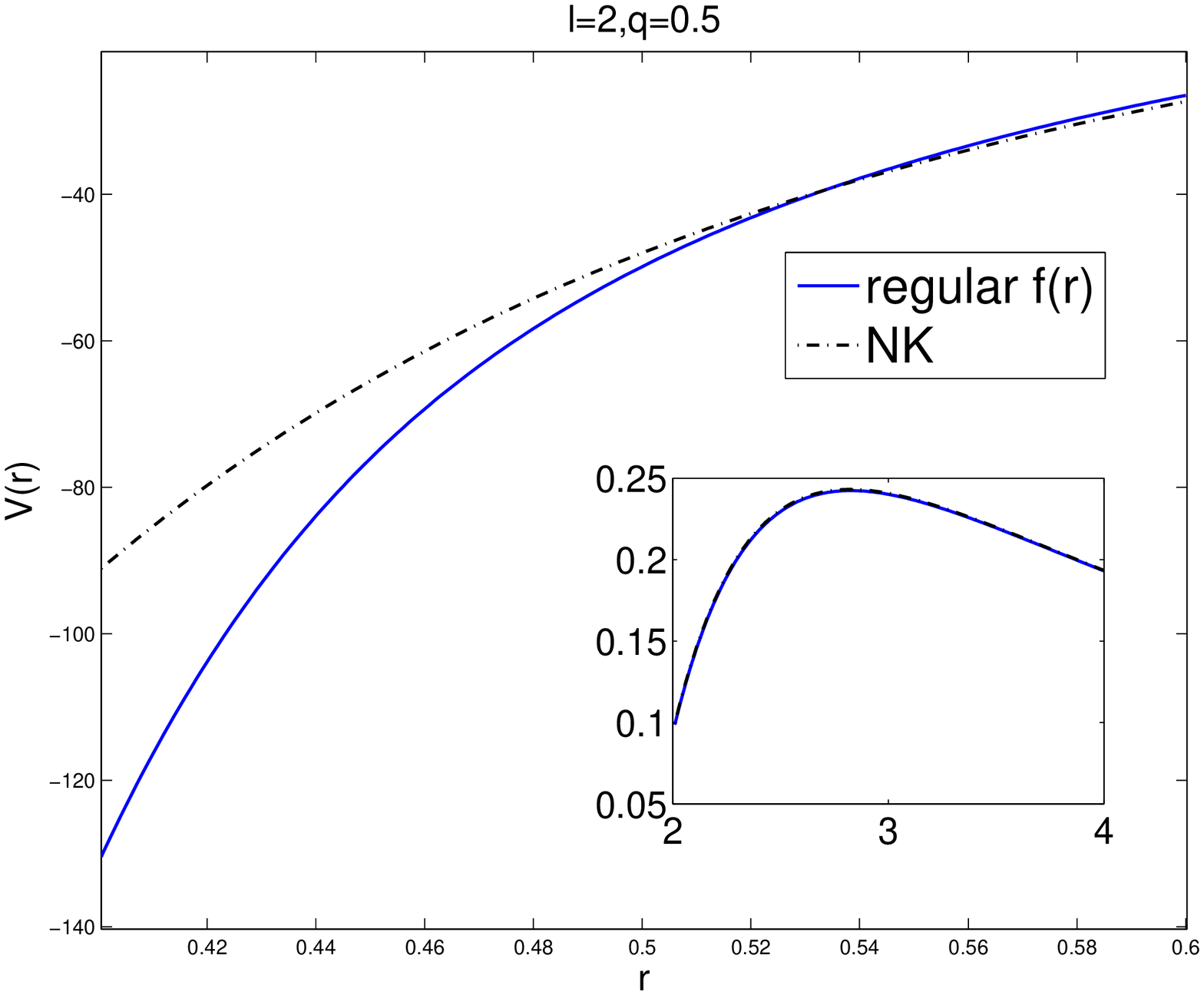}
\end{minipage}
} \caption{The potential function $V(r)$ of
the regular metric and RN BH ($M=1$). For regular BH, $q=q_{m}$, while
$q=q_{e}$ in RN condition} \label{fig:v(r)}
\end{figure}

As predicted, the potential behavior of the regular solution is quite similar to RN BH although the regular metric comes from NLED coupling to the Einstein equation while the RN solution is deduced from linear EM theory. FIG.\ref{fig:v(r)} displays the effective potential of the regular metric and the RN solution. We can find that their difference mainly concentrated in the area near $r=0$, and they always tend to be overlapped by each other once away from $r=0$. Therefore their QNMs frequencies would also be much similar, which can be reflected by FIG.\ref{fig:QNMreg-RN(r)} and Table~\ref{tab:expansion}.

\section{QNMs for the nonlinear electromagnetic field perturbation}\label{sec:regQNM}
The NLED EM perturbation is governed by the wave equation,
\begin{equation}
\left[\frac{d^{2}}{dr^{2}_{\ast}}+\omega^{2}-V(r,l)\right]\psi(t,r)=0,
\end{equation}
where $\psi(t,r)=\Phi(r)e^{-i\omega t}$. The solution satisfies the following boundary conditions:
\begin{enumerate}
\item Pure ingoing waves at the
event horizon $\Phi(r)\sim e^{-i\omega r_{\ast}}$,
$r_{\ast}\rightarrow -\infty$.
\item Pure outing waves at the
spatial infinity $\Phi(r)\sim e^{i\omega
r_{\ast}}$,$r_{\ast}\rightarrow \infty$.
\end{enumerate}
The WKB approximation can be used for effective potential that has the form of a potential barrier and takes constant values at the event horizon and spatial infinity. The method bases on matching the asymptotic WKB solutions at spatial infinity and the event horizon with a Taylor expansion near the top of the potential barrier through two turning points. Using the above potential functions, we calculate the QNMs frequencies through a 6$^{th}$order WKB approximation, which has been proven to be the most accurate method for finding the quasinormal spectrum with lower overtones~\cite{WKBusage}.

\begin{figure}
\centering \subfigure{
\begin{minipage}[t]{0.4\textwidth}
\includegraphics[width=1\textwidth]{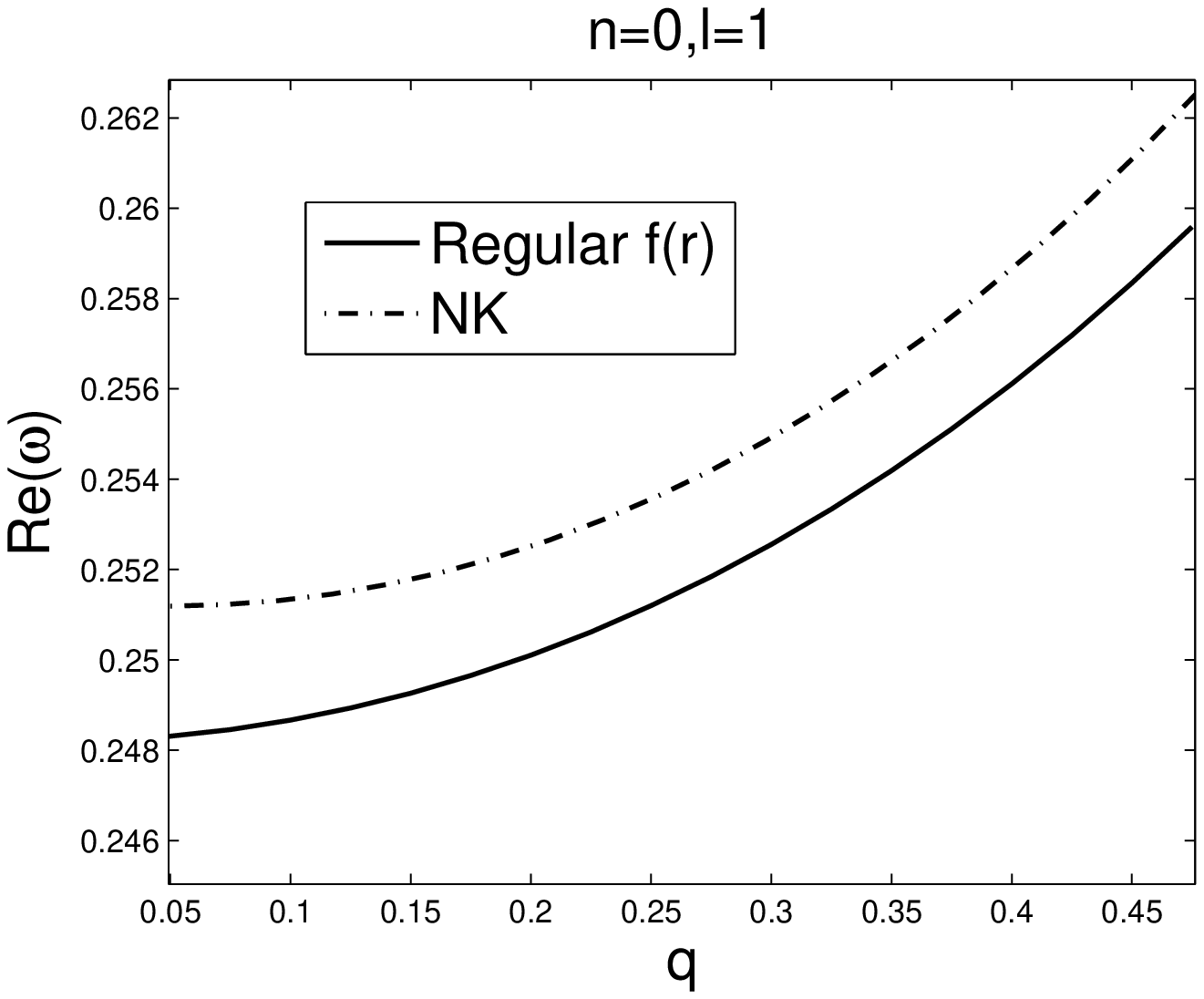}
\end{minipage}
} \subfigure{
\begin{minipage}[t]{0.4\textwidth}
\includegraphics[width=1\textwidth]{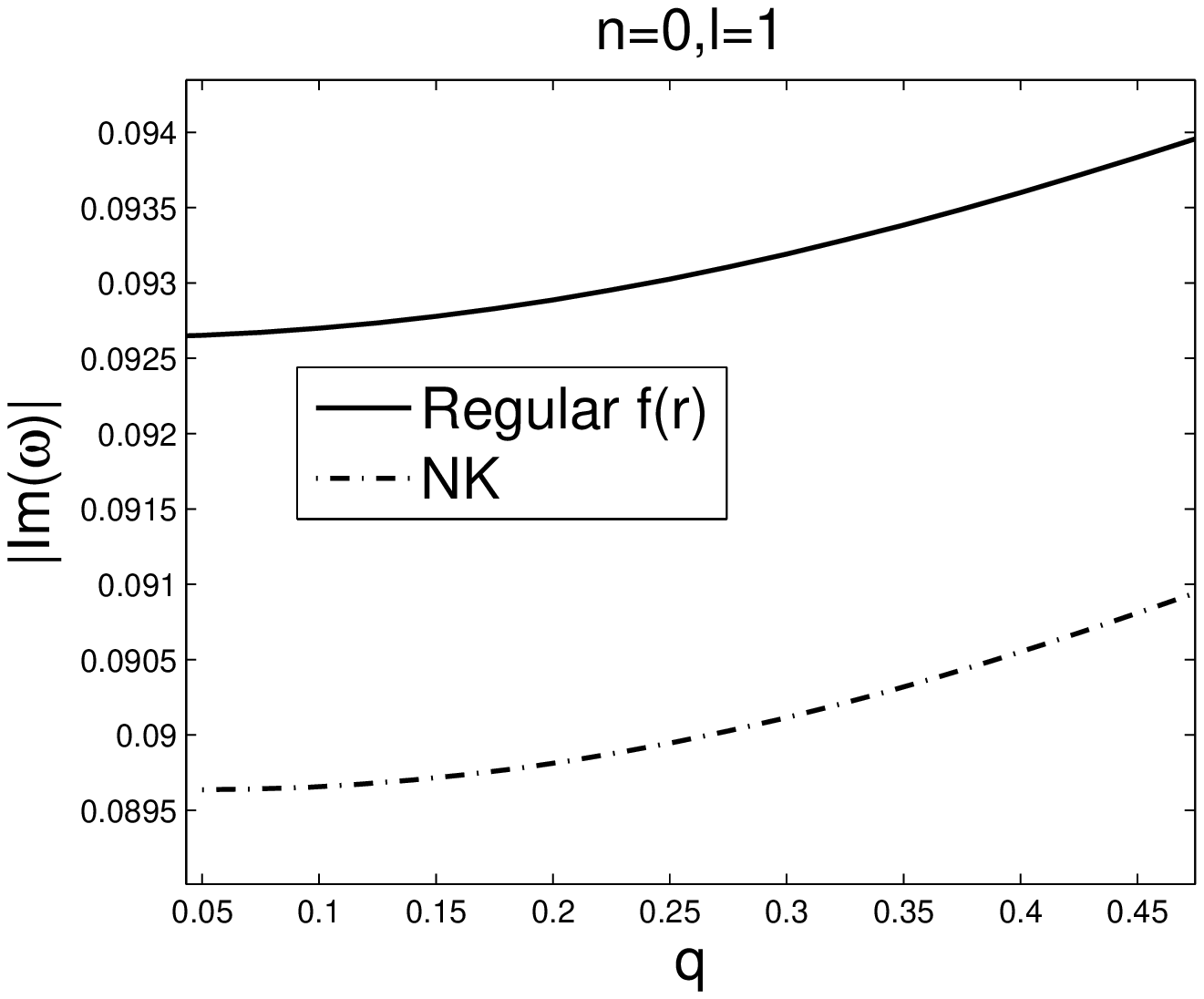} \\
\end{minipage}
}
\centering \subfigure{
\begin{minipage}[t]{0.4\textwidth}
\includegraphics[width=1\textwidth]{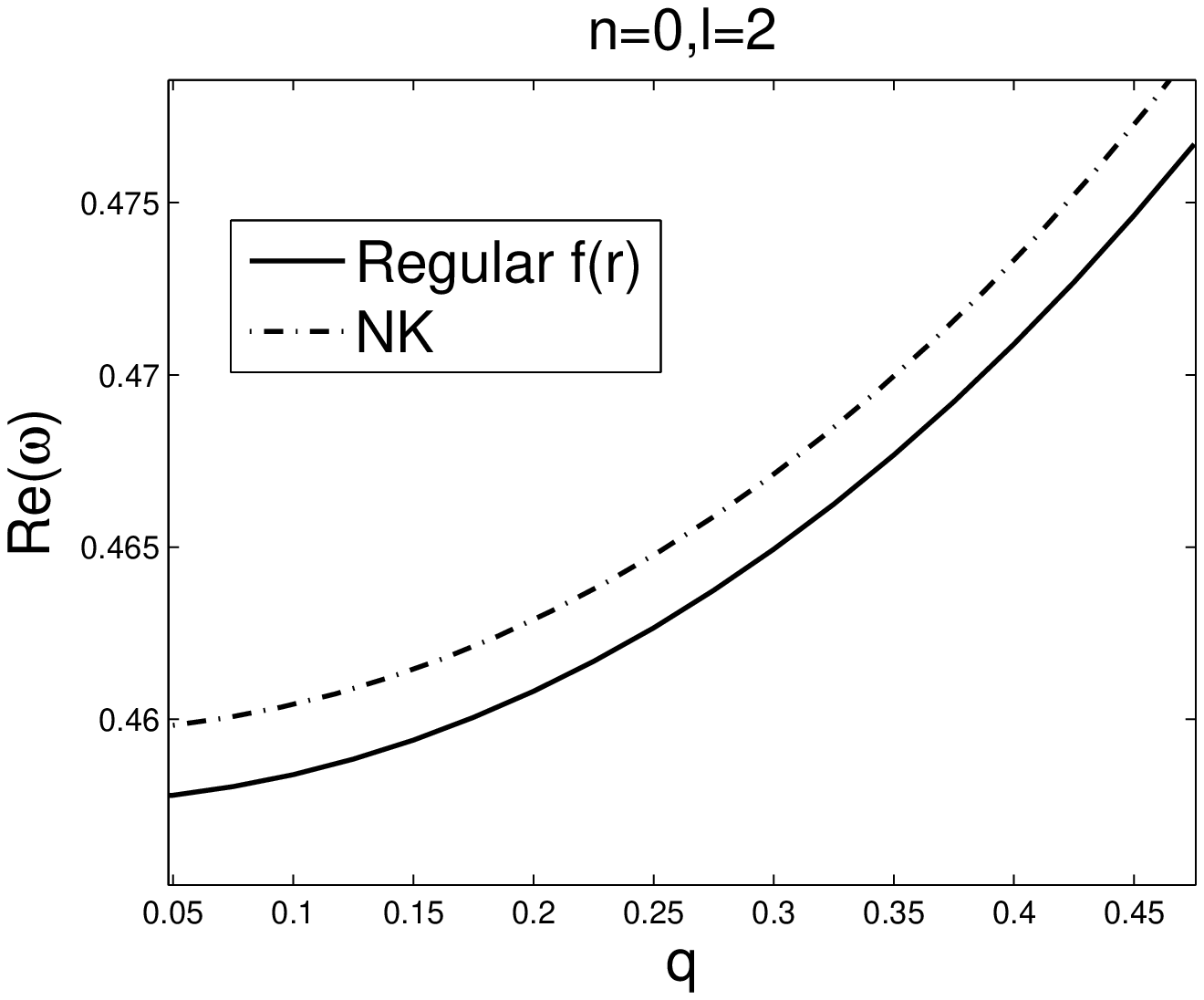}
\end{minipage}
} \subfigure{
\begin{minipage}[t]{0.4\textwidth}
\includegraphics[width=1\textwidth]{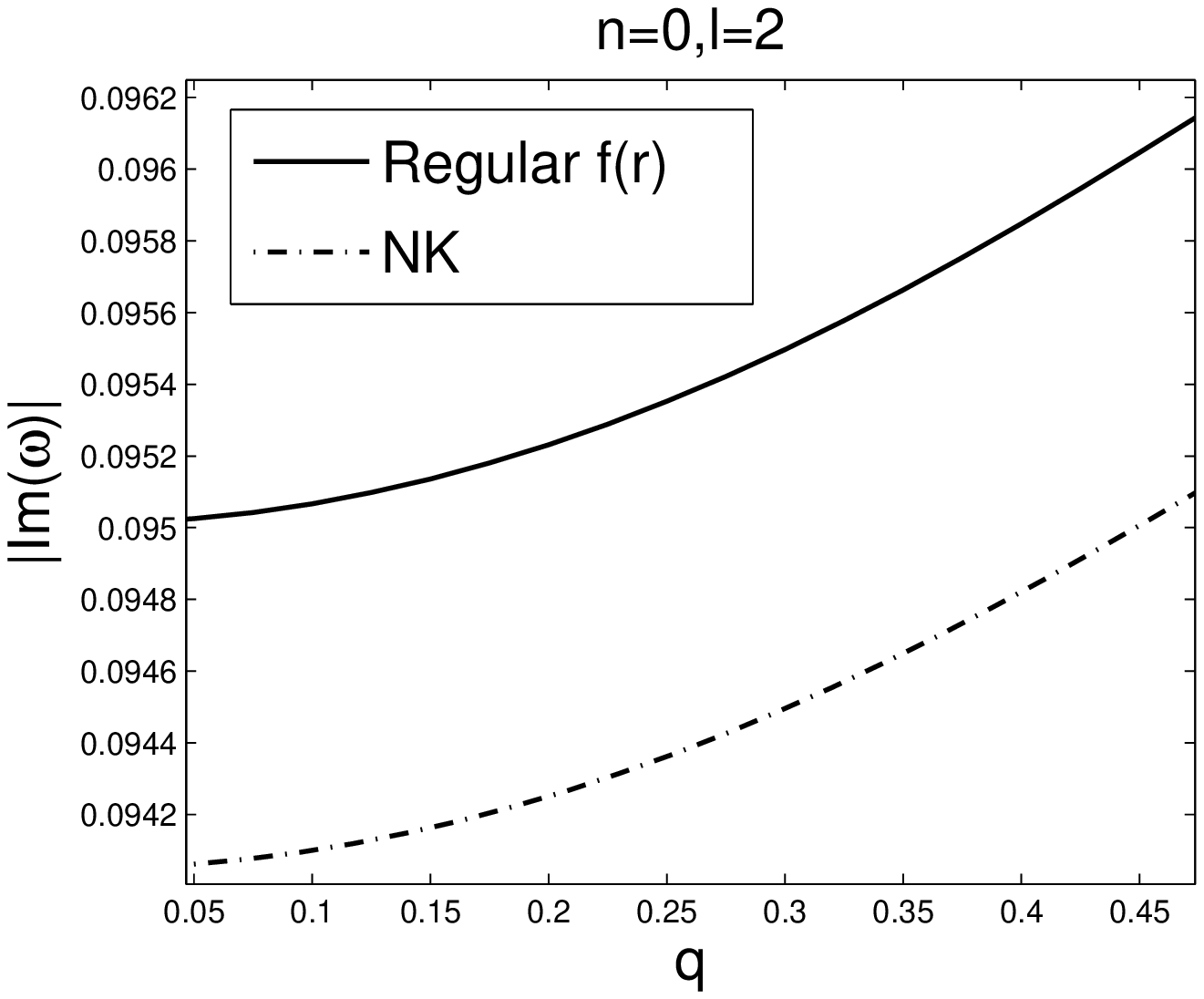} \\
\end{minipage}
}\caption{The comparison of QNMs frequencies from the regular BHs and RN BHs ($M=1$)}
\label{fig:QNMreg-RN(r)}
\end{figure}

Firstly, in order to illuminate the similarities of the QNMs from RN and such regular BH, it is necessary to compare their QNFs with some typical values. From FIG.\ref{fig:QNMreg-RN(r)}, it can be found that their discrepancy keeps in a small range, which is consistent with the results in Table \ref{tab:expansion} and more importantly varying the parameters their QNFs curves have the same shape. So the relationship between QNMs of the regular metric and parameters (such as $q$, $l$ and $n$) can stand for the cases of RN solution. Therefore in the following text, we only figure out the QNMs of the regular BH.

\begin{figure}
\centering \subfigure{
\begin{minipage}[t]{0.4\textwidth}
\includegraphics[width=1\textwidth]{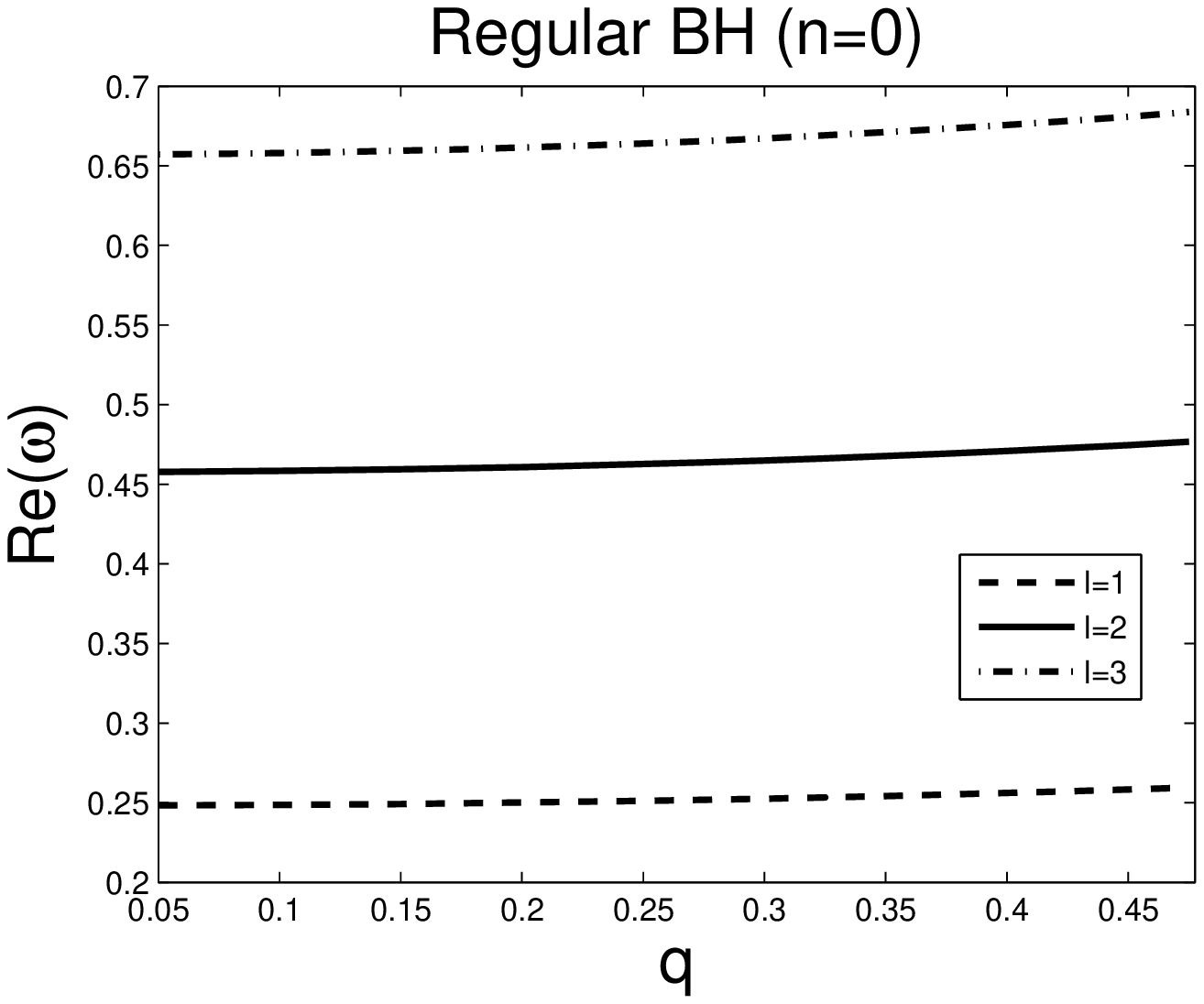}
\end{minipage}
} \subfigure{
\begin{minipage}[t]{0.4\textwidth}
\includegraphics[width=1\textwidth]{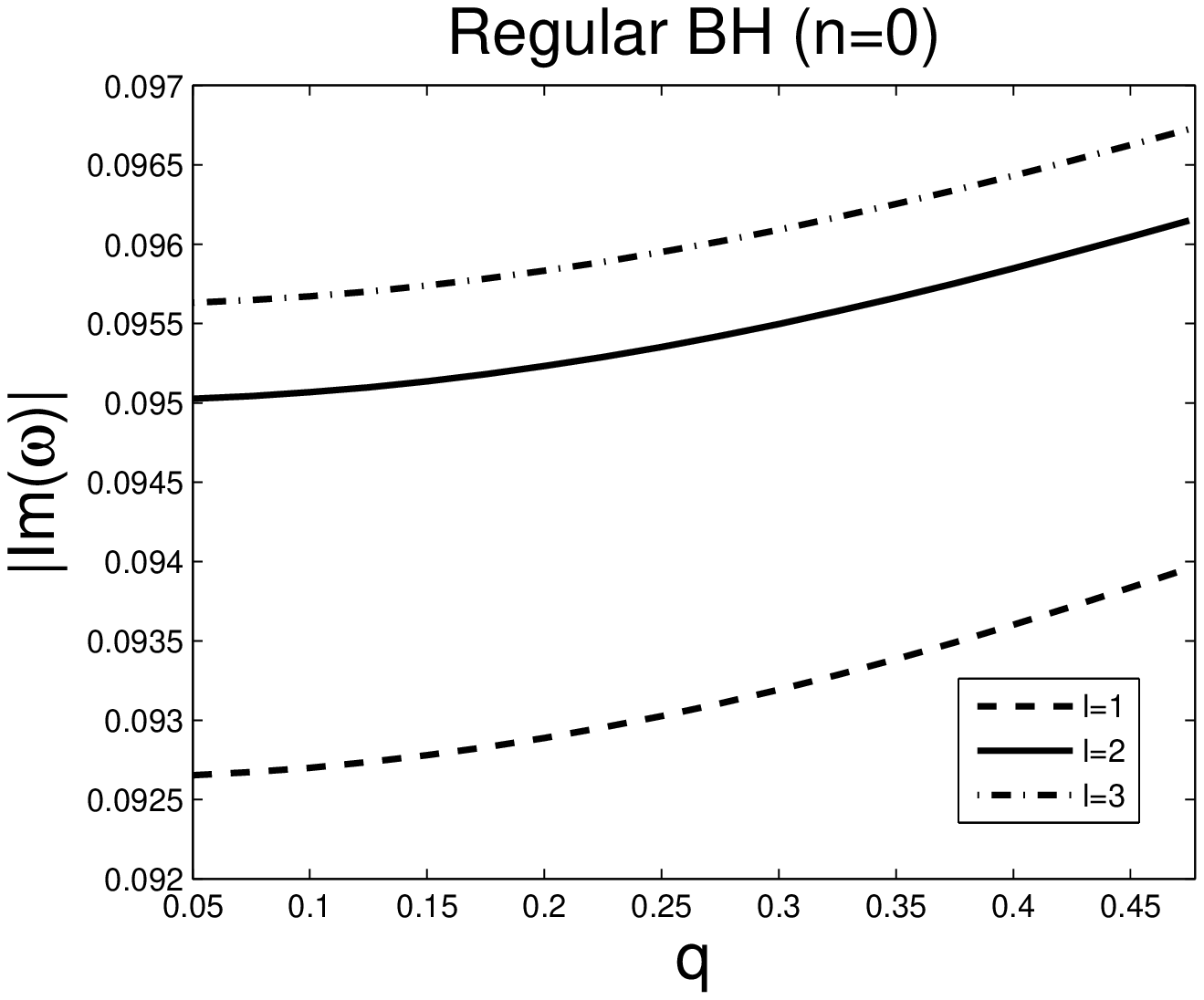} \\
\end{minipage}
}
\centering \subfigure{
\begin{minipage}[t]{0.4\textwidth}
\includegraphics[width=1\textwidth]{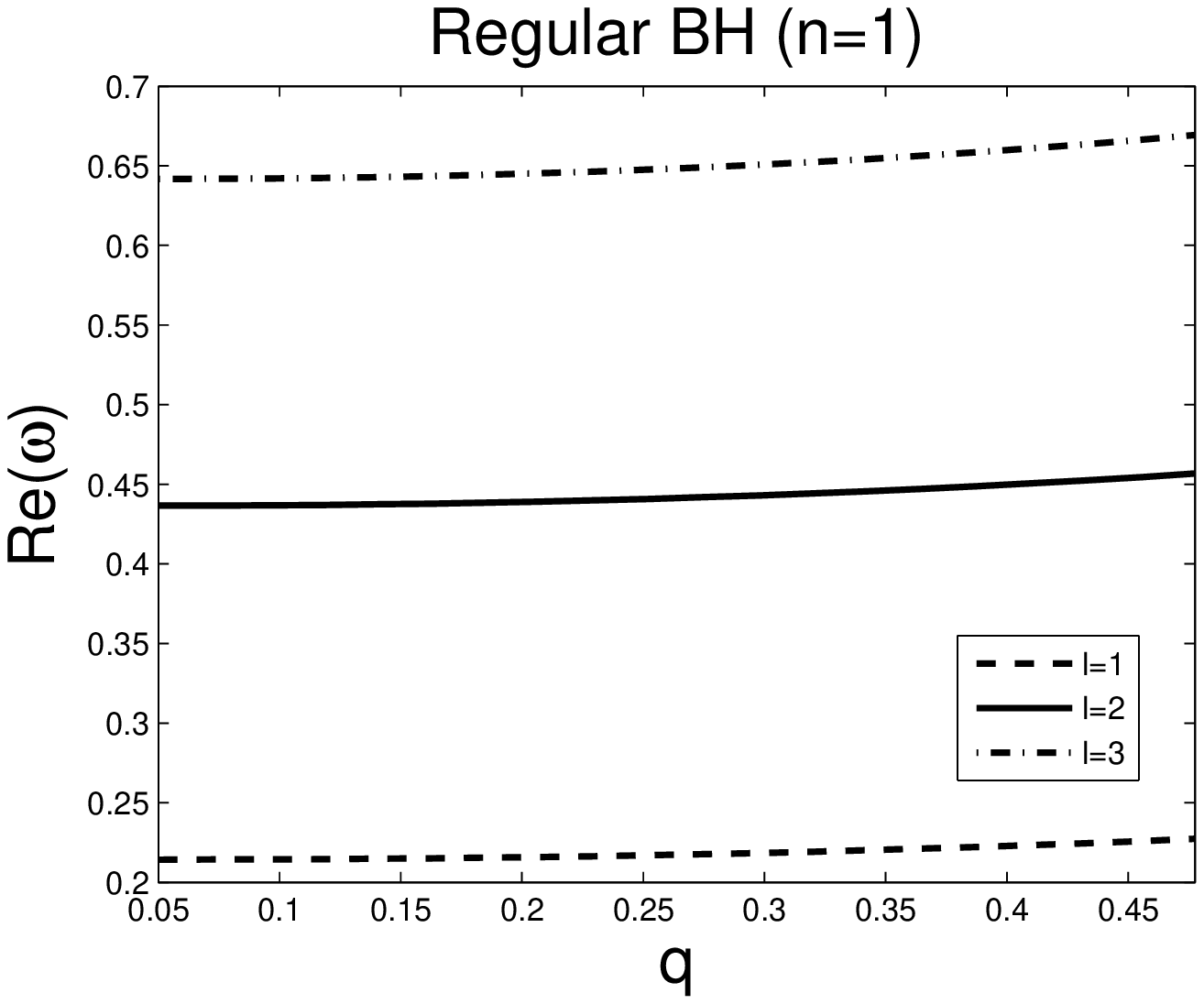}
\end{minipage}
} \subfigure{
\begin{minipage}[t]{0.4\textwidth}
\includegraphics[width=1\textwidth]{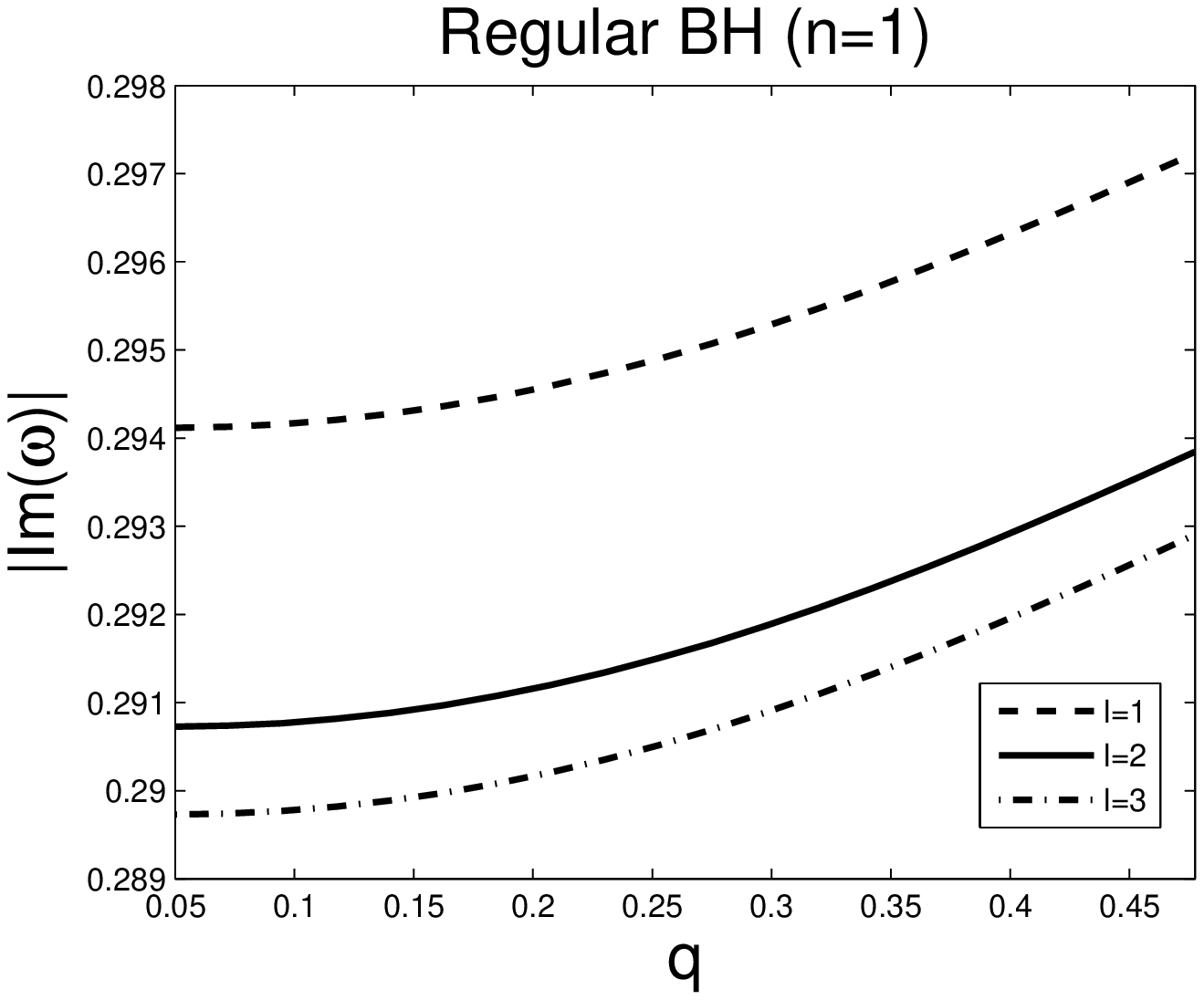} \\
\end{minipage}
}

\centering \subfigure{
\begin{minipage}[t]{0.4\textwidth}
\includegraphics[width=1\textwidth]{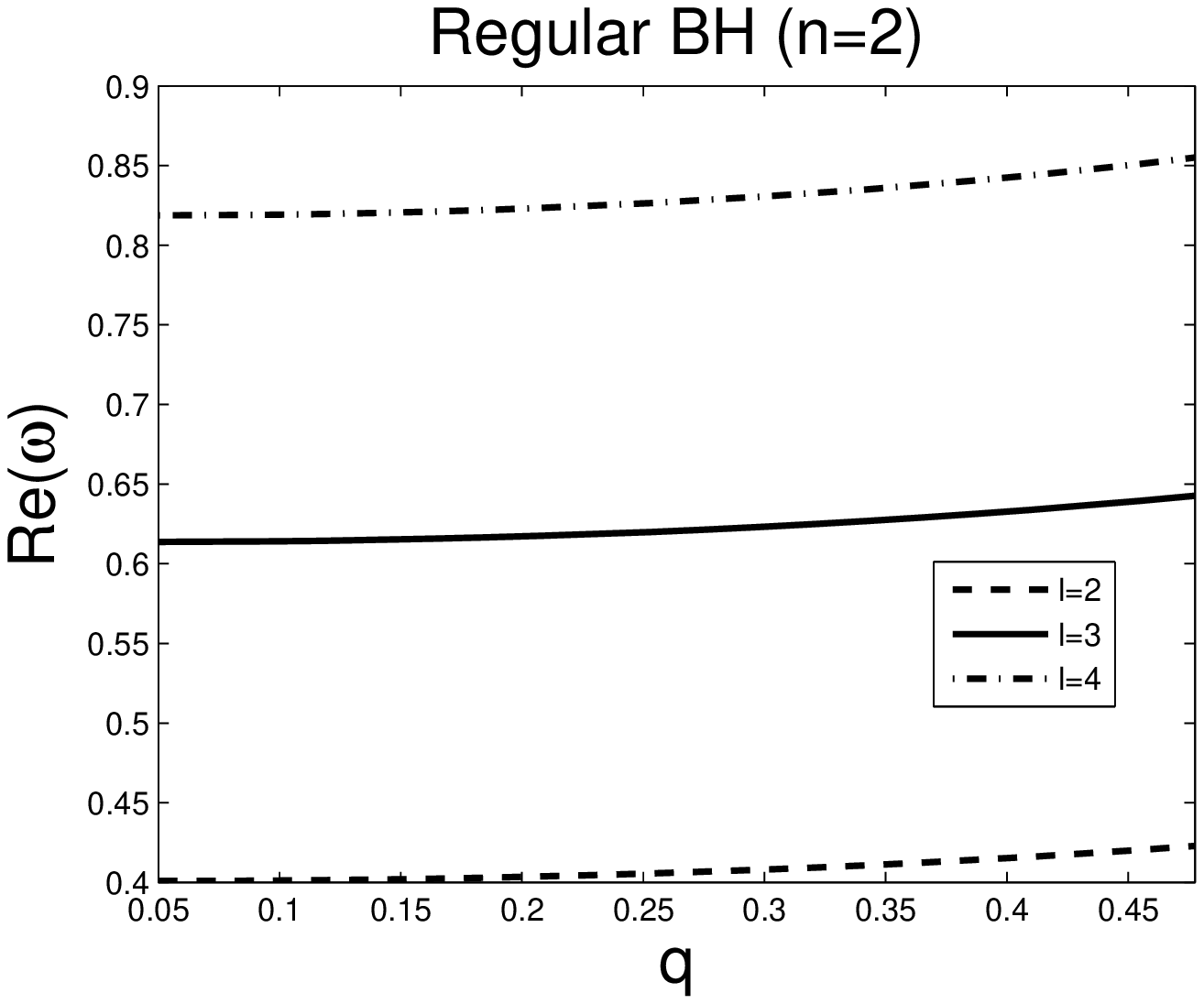}
\end{minipage}
} \subfigure{
\begin{minipage}[t]{0.4\textwidth}
\includegraphics[width=1\textwidth]{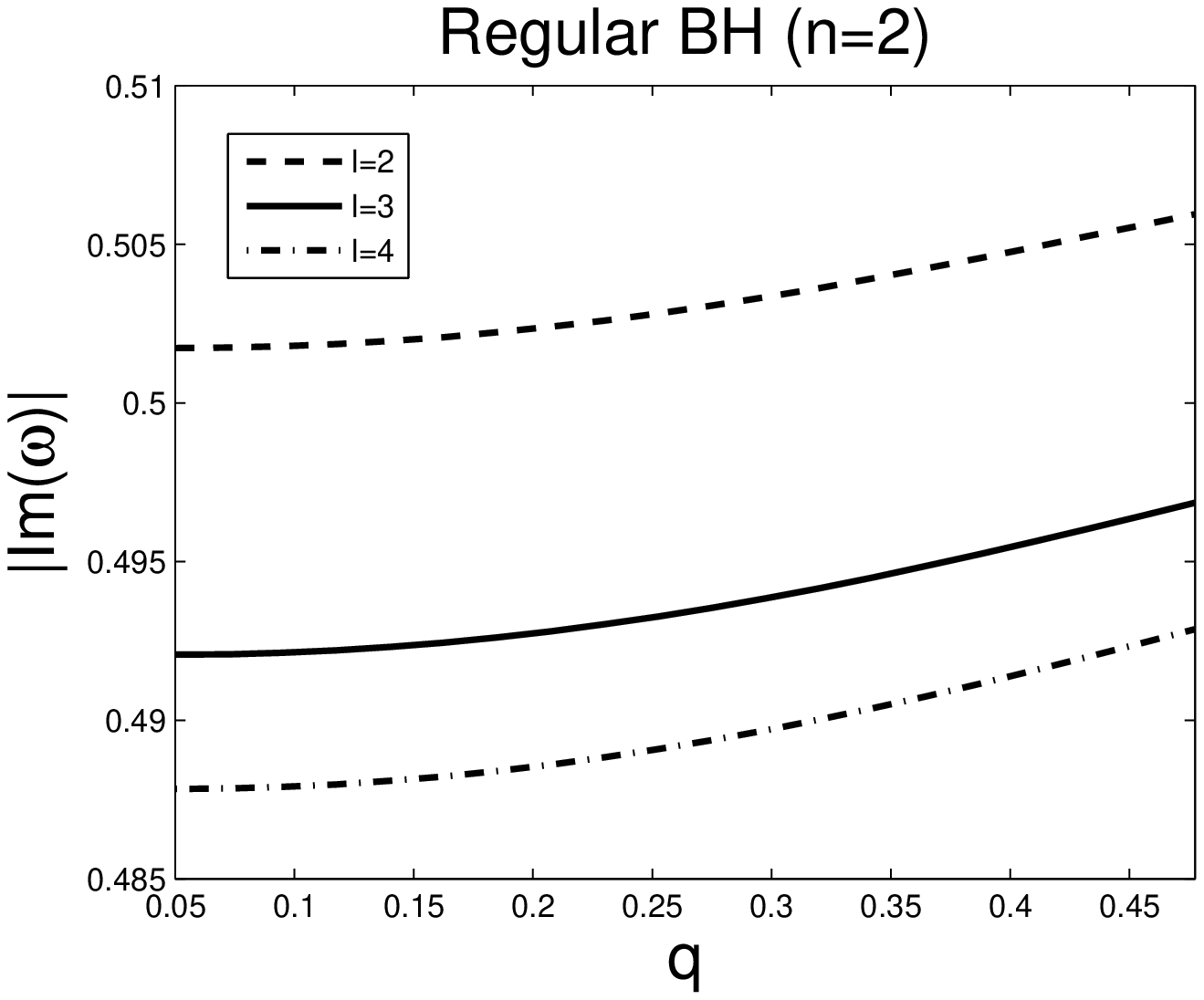} \\
\end{minipage}
}
\caption{The QNMs frequencies of the regular metric ($M=1$)}
\label{fig:QNMReg(r)}
\end{figure}

Secondly, we calculate the QNMs frequencies with $n=0,1,2$, and for each overtone $n$ we choose $l=n+1,n+2,n+3$ to determine how the magnetic charge impacts on the QNMs. FIG.\ref{fig:QNMReg(r)} shows the shapes of QNMs with different $n$ and $l$. For the real part of $\omega$, it increases with larger $q$ and $l$, however decrease slightly with higher overtone. This means the perturbation in the fundamental mode (i.e.,$n=0$) with larger $q$ and $l$ leads to more intense QNMs oscillation. For the imaginary part of $\omega$, we can find that 1) given $n,l$, $|\rm{Im}(\omega)|$ increases with stronger charge. 2) given $q$, in $n=0$ case the higher $l$ enhances $|\rm{Im}(\omega)|$ while in $n=1,2$ cases $l$ has an inverse effect on $|\rm{Im}(\omega)|$. This can be seen as a property of fundamental QNMs oscillation. Overall, for such regular BH the fundament mode with lower charge and angular momentum number plays the dominant role in QNMs oscillation since it exists in the longest time.

Next, we employ the expansion method to evaluate the QNMs frequencies. It has been known for many years that QNMs are intimately linked to the existence and properties of unstable null orbits, and then Dolan and Ottewill developed a simple method for determining the QNMs frequency $\omega$, which can be expanded in inverse powers of $\hat{l}=l+1/2$~\cite{eikonal1}.

According to the expansion method proposed by Dolan and Ottewill, the radial function $\Phi(r)$ can be redefined as
\begin{equation}
\label{eikonal1} \Phi(r)=\upsilon(r)e^{\int^{r_{*}}\alpha(r)dr_{*}},
\end{equation}
where $\alpha(r)=i\omega b_{c}k_{c}(r)$ and
\begin{equation}
k_{c}(r)=(r-r_{c})\sqrt{\frac{k^{2}(r,b_{c})}{(r-r_{c})^{2}}},~~~~~~
k^{2}(r,b)=\frac{1}{b^{2}}-\frac{f(r)}{r^{2}},
\end{equation}
with the condition
\begin{equation}
k^{2}(r_{c},b_{c})=\frac{\partial k^{2}(r,b_{c})}{\partial
r}\Big|_{r=r_{c}}=0,
\end{equation}
yielding the wave equation as
\begin{equation}
\label{eikonal4}
\frac{d^{2}\upsilon(r)}{dr_{*}}+2\alpha(r)f(r)f'(r)+[\omega^{2}+\alpha^{2}(r)-V(r)+f(r)\alpha'(r)]\upsilon(r)=0.
\end{equation}
For the fundamental mode $n=0$,  $\omega$ and $\upsilon(r)$ can
be expanded as
 \bqn
\label{eikonal5}
\omega=\sum\limits_{i=-1}^{\infty}\left(\frac{a_i}{\hat{l}^i}\right),~~~~~~
\ln v(r)=\sum\limits_{i=0}^{\infty}\left[\hat{l}^{-i}S_i(r)\right]. \eqn

So the key advantage of the expansion method over many other approaches
is that it furnishes us with simple approximations for the wave function (i.e., eq.(\ref{eikonal1})) and it is more accurate in the case $l>n$. In order to illustrate the accuracy of the expansion method, we calculate some QNMs frequencies with specific parameters and angular momentum numbers for the regular and RN metrics (cf. Table~\ref{tab:expansion}). The results show good agreement with those from the WKB approximation.
\begin{table}[!h]
 \tabcolsep 0pt \caption{QNMs frequencies evaluated by expansion method ($n=0,M=1$ ignoring $L^{-3}$ term)}\label{tab:expansion}
 \vspace*{-12pt}
\begin{center}
\def\temptablewidth{1\textwidth}
{\rule{\temptablewidth}{1pt}}
\begin{tabular*}{\temptablewidth}{@{\extracolsep{\fill}}cccc}
$q$ & $l$ & $\text{Regular}~\text{metric}$ &$\text{RN}~\text{metric}$ \\
\hline
      $ $ & $1$ &$ 0.250538-0.0930387i$  &$0.252657-0.0897465i$ \\
      $0.1$ &$2$ &$ 0.458758-0.0951121i $  &$0.460029-0.0939269i$ \\
      $ $ & $3$ &$0.65815-0.0956833i $  &$0.659059-0.0950786i$ \\
      \hline
      $ $ & $1$ &$ 0.251969-0.093213i$  &$0.254115-0.0898996i$ \\
      $0.2$ &$2$ &$ 0.461174-0.0952761i$  &$0.462472-0.0940833i$ \\
      $ $ & $3$ &$0.661545-0.0958445i $  &$0.662484-0.0952359i$ \\
      \hline
      $ $ & $1$ &$0.254405-0.0934975i$  &$0.256624-0.0901488i$ \\
      $0.3$ &$2$ &$0.465284-0.0955432i$  &$0.46667-0.0943376i$ \\
      $ $ & $3$ &$0.667317-0.0961069i$  &$0.668367-0.0954916i$ \\
      \hline
      $ $ & $1$ &$0.257929-0.0938805i$  &$0.260315-0.0904819i$ \\
      $0.4$ &$2$ &$0.471218-0.0959013i$  &$0.472834-0.0946768i$ \\
      $ $ & $3$ &$0.675649-0.096458i$  &$0.677001-0.0958326i$ \\
      \hline
      $ $ & $1$ &$0.262671-0.0943383i $  &$0.265402-0.0908729i$ \\
      $0.5$ &$2$ &$0.479186-0.096326i$  &$0.481308-0.095074i$ \\
      $ $ & $3$ &$0.686829-0.0968736i$  &$0.688863-0.0962315i$ \\
       \end{tabular*}
       {\rule{\temptablewidth}{1pt}}
       \end{center}
\end{table}

Finally in order to determine the QNMs oscillation shape, we adopt finite difference method to study the dynamical
evolution of the NLED field perturbation in the time
domain and examine the stability of the regular black
hole.

\begin{figure}
\centerline{\includegraphics[height=4cm]{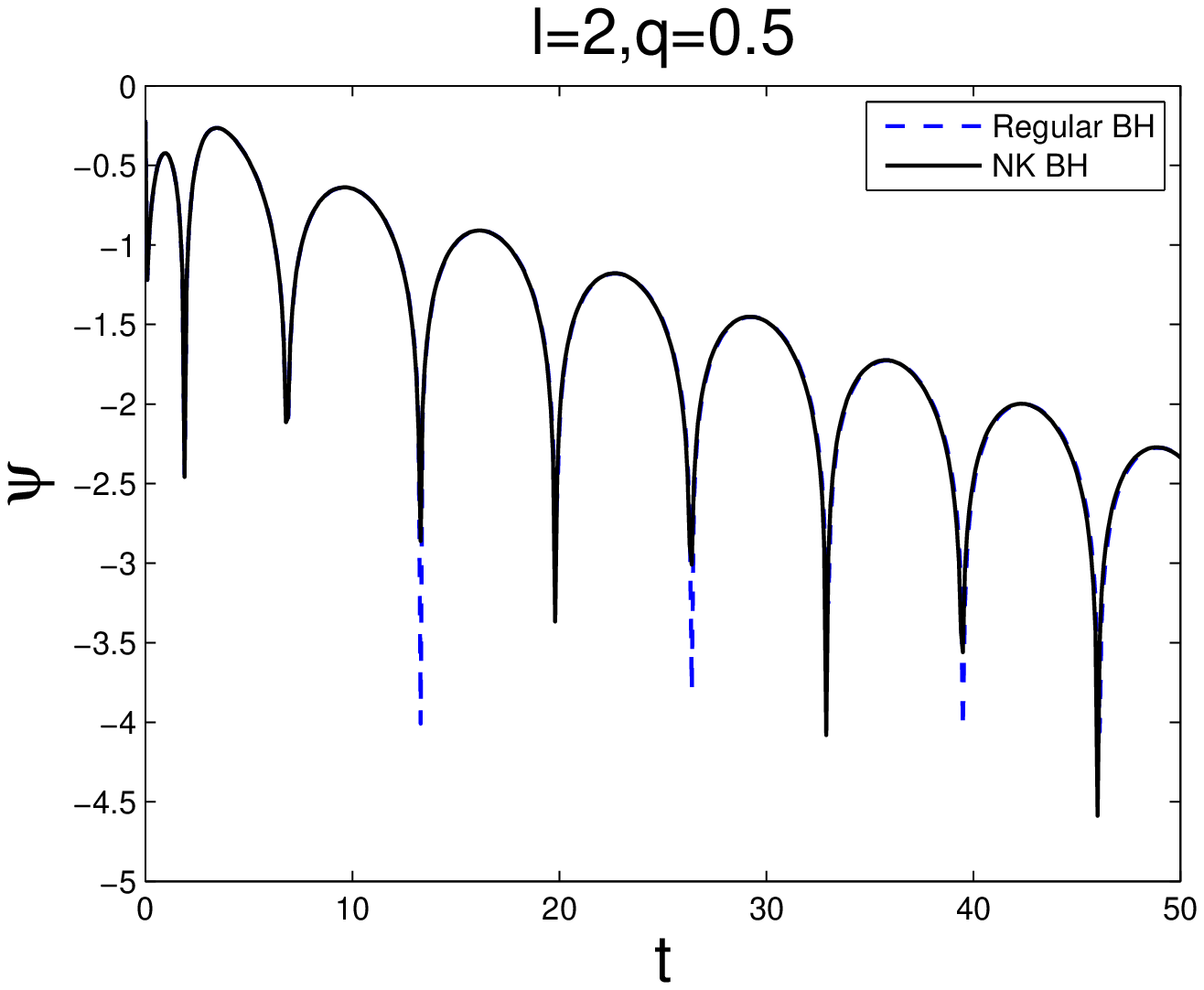}\includegraphics[height=4cm]{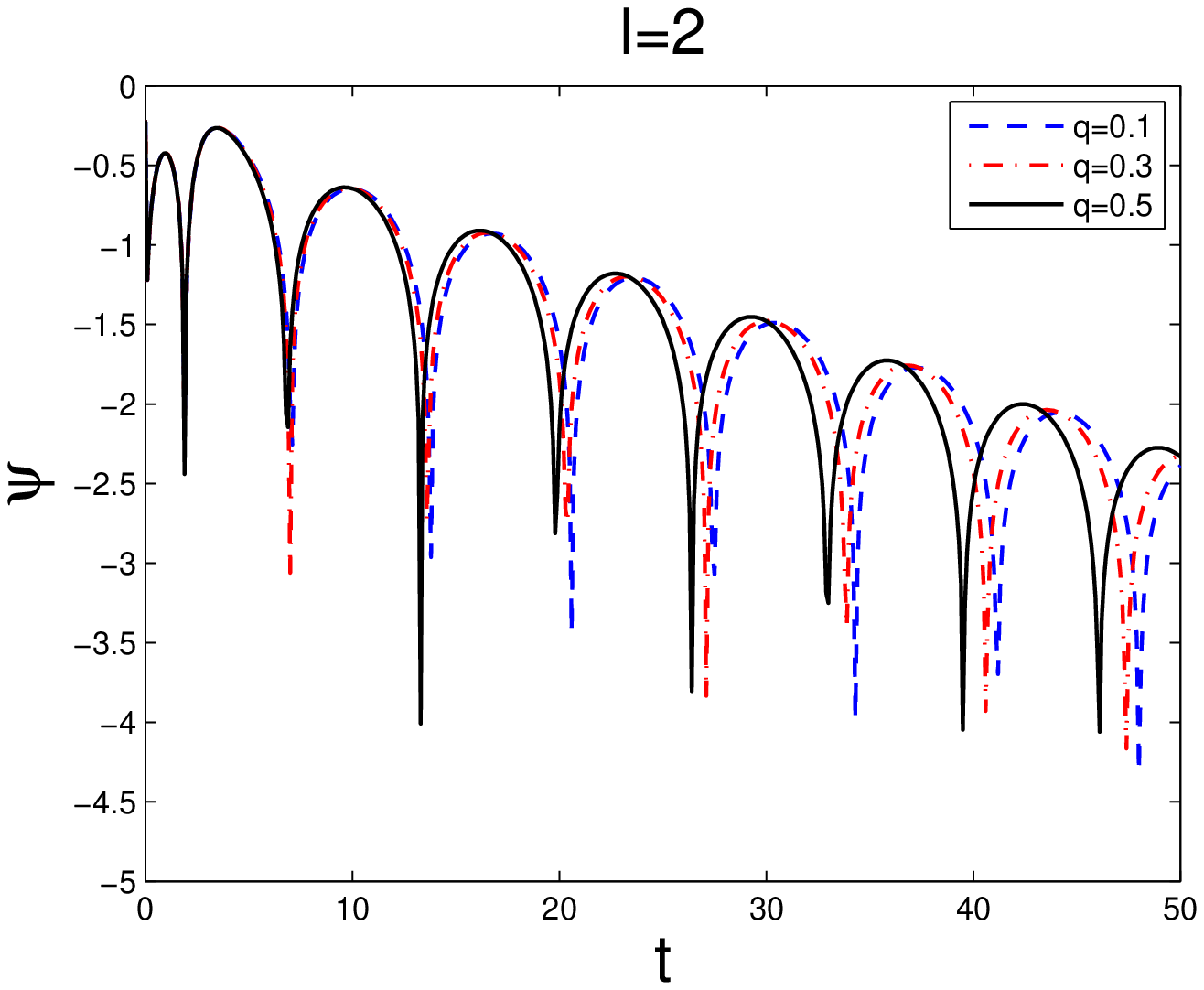}\includegraphics[height=4cm]{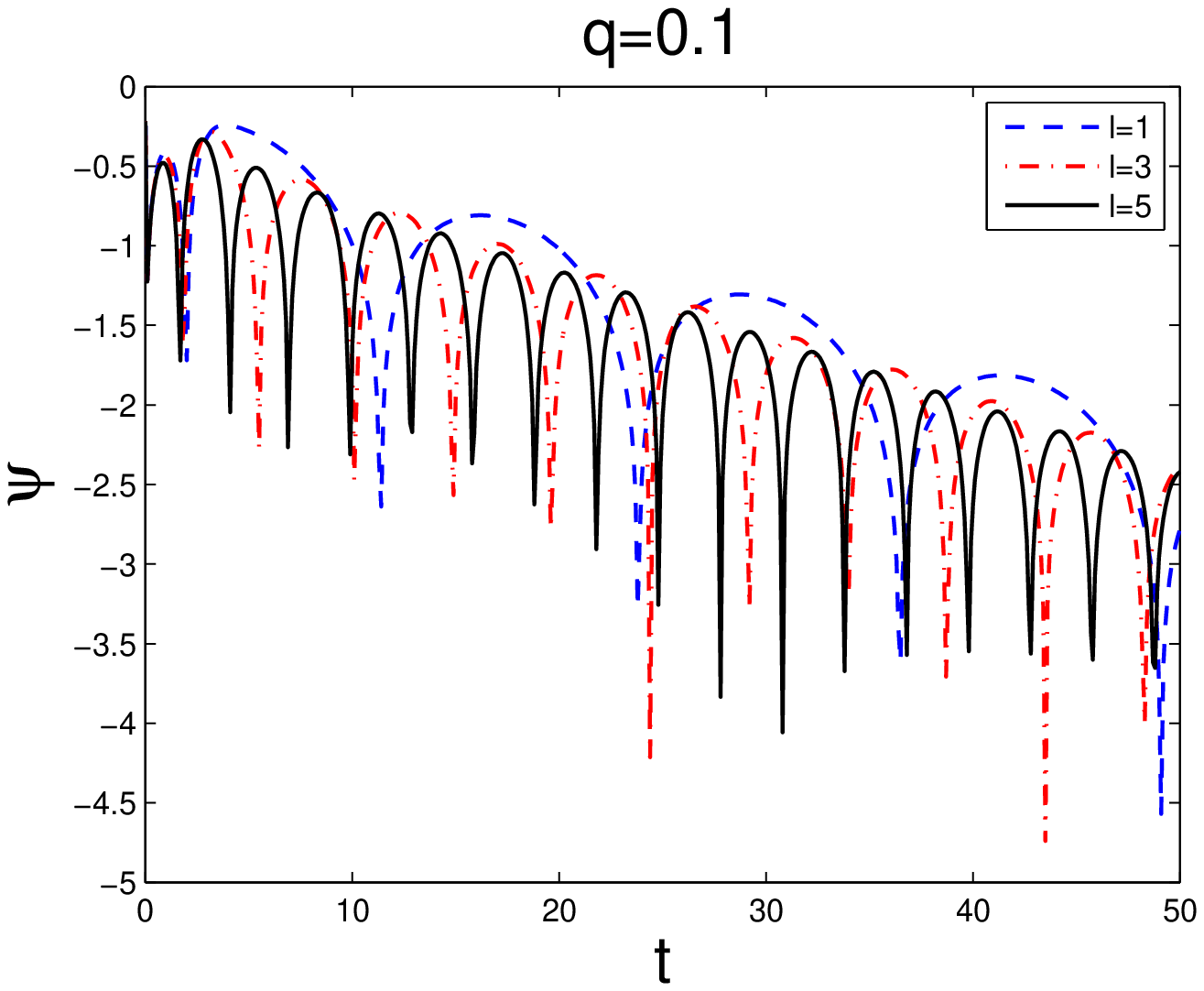}}
\caption{The dynamical evolution of NLED EM field in the background of the regular black
hole spacetime ($M=1$). The
constants in the gaussian pulse $\nu_{c}=1$ and $\sigma=1$.}
\label{Fig:finitReg}
\end{figure}

Firstly,we re-write the wave equation in terms of the variables $\mu=t-r_{\ast}$
and $\nu=t+r_{\ast}$ ($dr_{\ast}=dr/f(r)$):
\begin{equation}
4\frac{\partial^{2}\Psi}{\partial \mu\partial \nu}+V(r)\Psi=0.
\end{equation}
This two-dimensional wave equation can be integrated numerically by
using the finite difference method suggested in
~\cite{Finit1,Finit2}.
 It can be discretized as
 \bqn
 \lb{newadd}
\psi\left(\mu+\delta\mu,\nu+\delta\nu\right)&=&\psi\left(\mu,\nu+\delta\nu\right)+\psi\left(\mu+\delta\mu,\nu\right)-\psi\left(\mu,\nu\right)\nb\\
&&-\delta\mu\delta\nu
V\left(\frac{2\nu-2\mu+\delta\nu-\delta\mu}{4}\right)\frac{\psi\left(\mu+\delta\mu,\nu\right)+\psi\left(\mu,\nu+\delta\nu\right)}{8}+{\cal
O}\left(\epsilon^{4}\right),
 \eqn
where $\epsilon$ is an overall grid scale factor
(i.e.,$\delta\mu\sim\delta\nu\sim\epsilon$). Set a boundary
condition $\psi(\mu,\nu=\nu_{0})=0$ and suppose an initial perturbation as a gaussian pulse, centered on $\nu_{c}$ and with width
$\sigma$ on $\mu=\mu_{0}$ as
\begin{equation}
\psi(\mu=\mu_{0},\nu)=e^{-\frac{(\nu-\nu_{c})^{2}}{2\sigma^{2}}}.
\end{equation}
The above equations indicate the key steps for numerical calculation of the field dynamical evolution. FIG.\ref{Fig:finitReg} is just the dynamical evolution of the NLED EM field in the background of the regular space-time. As expected, the QNMs oscillation of RN and the regular solution are almost the same (see the left subplot). The oscillation frequency increases with $q$ and $l$, and decay speed becomes slightly faster with larger $q$ and $l$, which are in accordance with the results of WKB method.
\section{Near-extreme cases}\label{sec:extremecondition}
\begin{figure}
\centering \subfigure{
\begin{minipage}[t]{0.4\textwidth}
\includegraphics[width=1\textwidth]{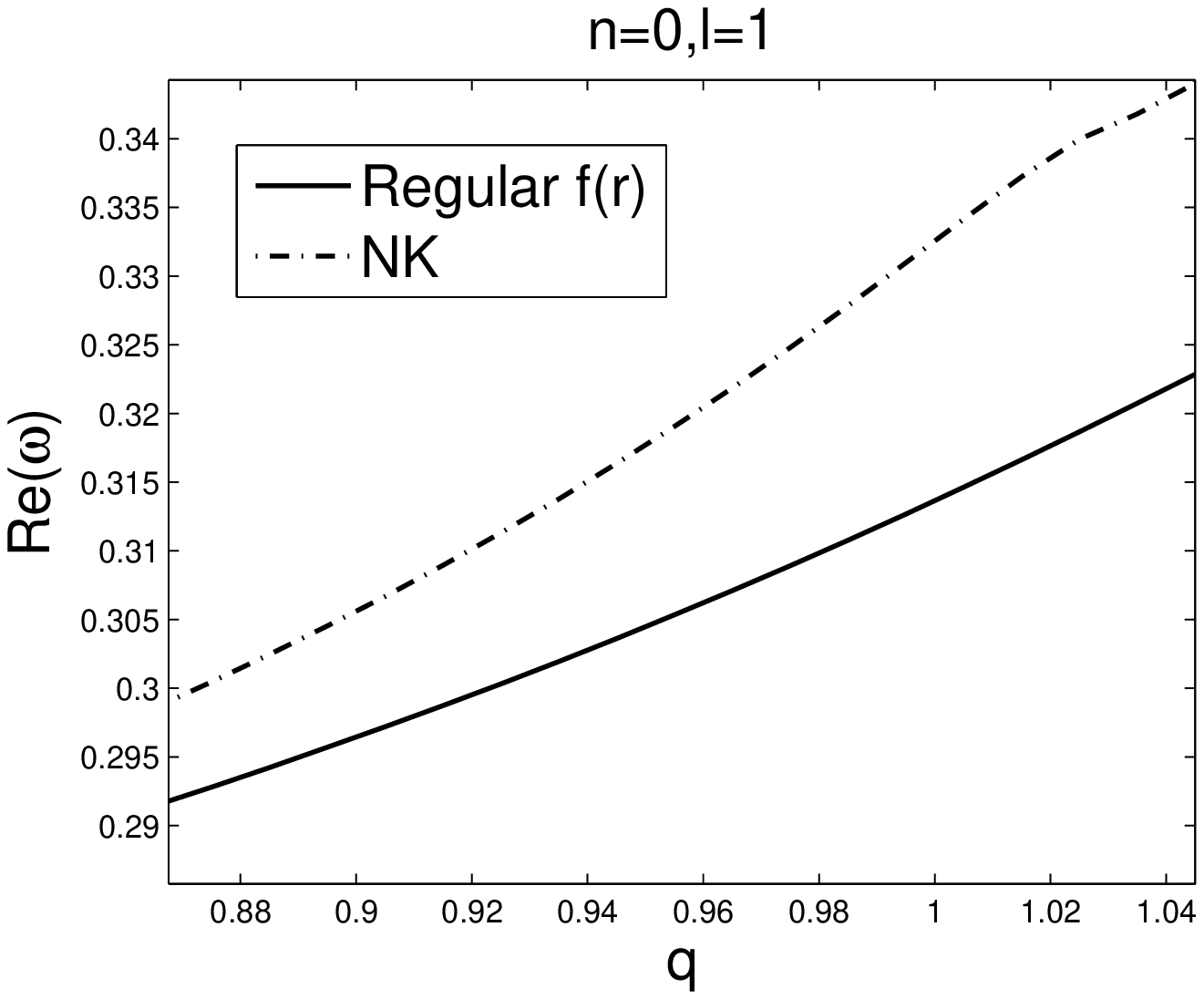}
\end{minipage}
} \subfigure{
\begin{minipage}[t]{0.4\textwidth}
\includegraphics[width=1\textwidth]{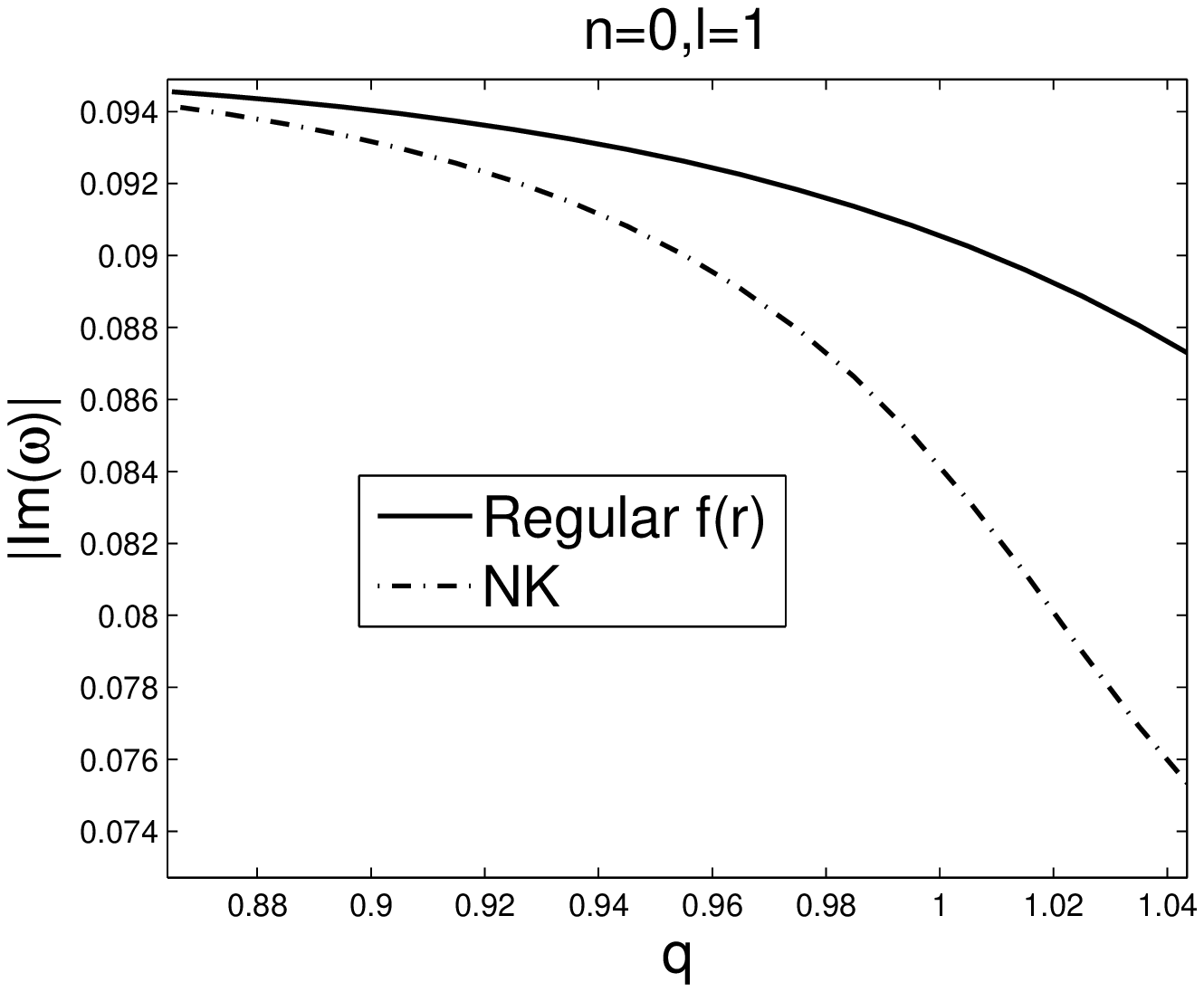} \\
\end{minipage}
}
\centering \subfigure{
\begin{minipage}[t]{0.4\textwidth}
\includegraphics[width=1\textwidth]{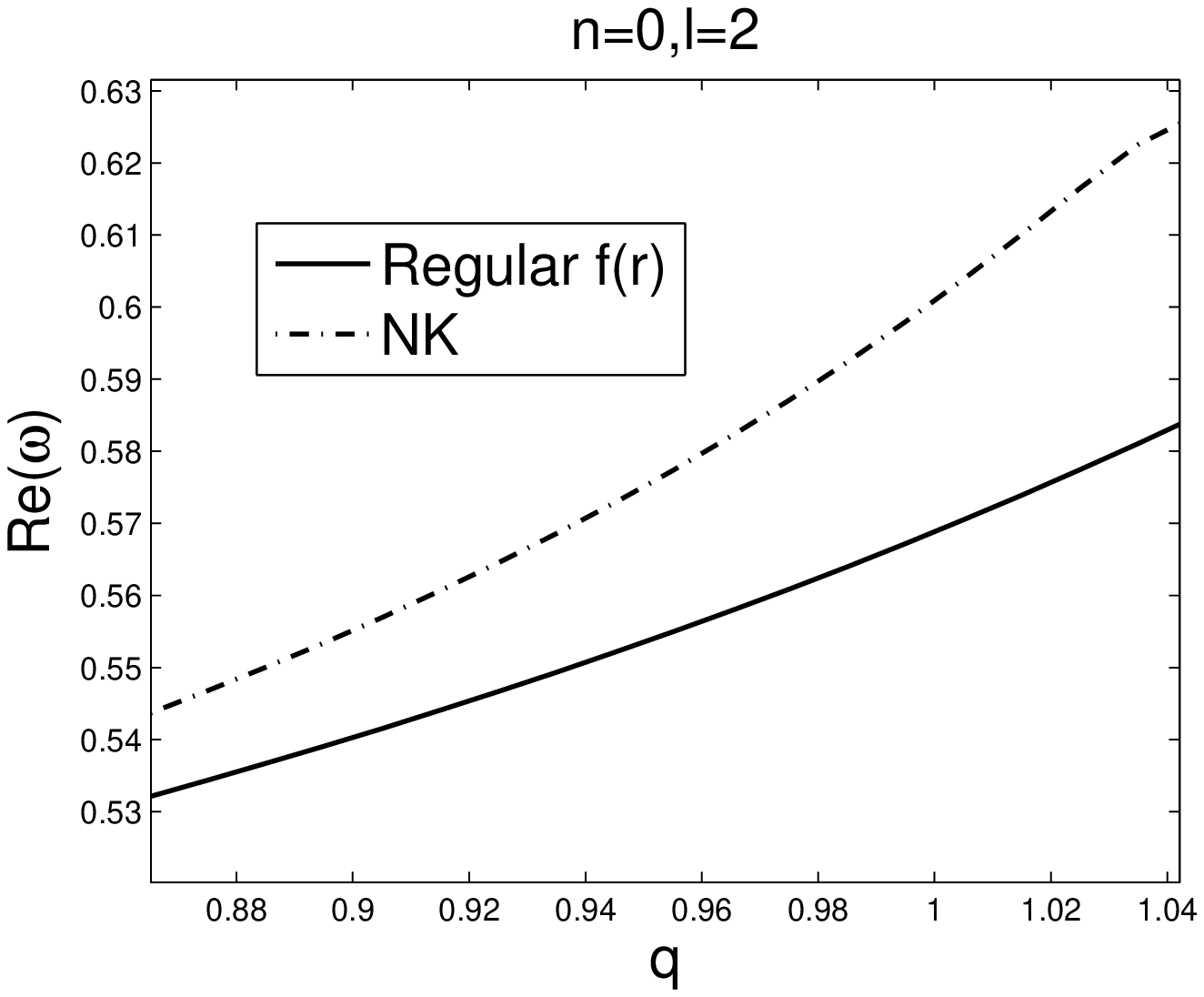}
\end{minipage}
} \subfigure{
\begin{minipage}[t]{0.4\textwidth}
\includegraphics[width=1\textwidth]{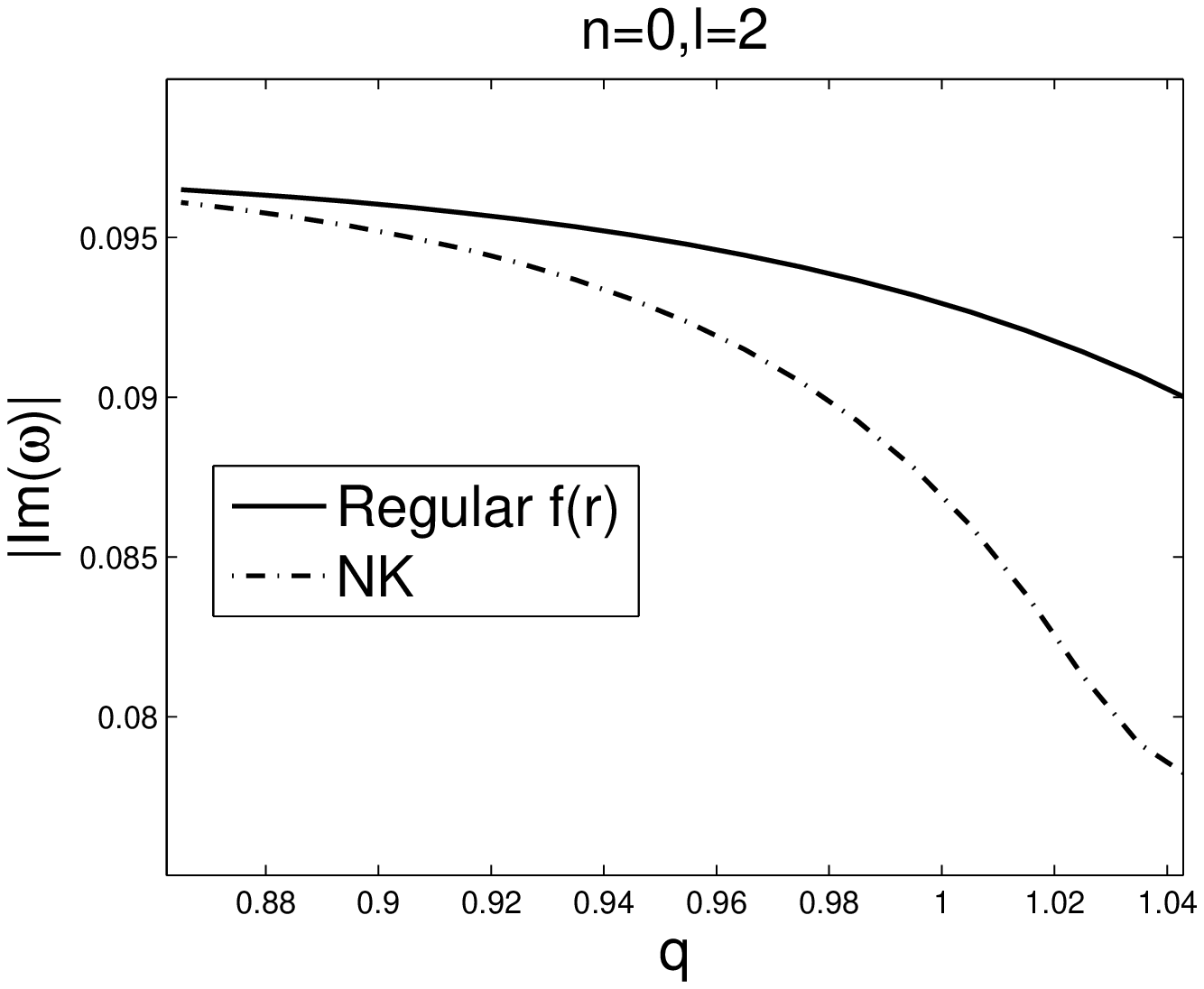} \\
\end{minipage}
}
\caption{The comparison of QNMs frequencies of the regular metric and RN BH in near-extreme cases}
\label{fig:QNMexRN-reg(r)}
\end{figure}

In this section, we consider the near-extreme cases around the extreme condition. When the inner horizon $r_{+}$ coincides with the outer horizon $r_{0}$, the extreme condition occurs. In order to maintain cosmic censorship~\cite{Cosmic1,Cosmic2}, the mass $M$ and $q$ should be confined by certain relationship in order to keep $r_{+}\leq r_{0}$.
So we have
\begin{equation}
\label{ex1}
f(r)=1-\frac{2M}{r}\left(1-\tanh\frac{|q|^{2}}{2Mr}\right)=0,
\end{equation}
\begin{equation}
\label{ex2}
f'(r)|_{r=r_{0}}=0.
\end{equation}
The second equation comes from the requirement that the temperature of a BH on the extreme condition should be zero. Through Eq.(\ref{ex1}) and Eq.(\ref{ex2}), the location of the event horizon can be expressed explicitly as $r_{0} = 4Mq'/(q'-W(-q'\exp(q')))$, where $W(x)$ is Lambert function and $q'=q^{2}/(2M)^{2}$. So we find one threshold quantity as $(q_{c}\simeq 1.0554,r_{0}\simeq 0.8712)$. Then we plan to analyze the QNMs around the threshold charge $q_{c}$. Due to the special function $\text{tanh}$ in $f(r)$, the potential function has a complicated form causing endless computation and inaccurate 6$^{th}$ order WKB results. Therefore we employ a Taylor expansion around $q_{c}$ and calculate the QNMs by 3$^{rd}$ order WKB approximation. Given the common parameters, FIG.\ref{fig:QNMexRN-reg(r)} compares the QNMs of the regular BH with that of RN BH around extreme condition, and indicates that in the near-extreme cases the regular BH after the NLED EM perturbation tends to stability faster than the RN BH.
\begin{figure}
\centering \subfigure{
\begin{minipage}[t]{0.4\textwidth}
\includegraphics[width=1\textwidth]{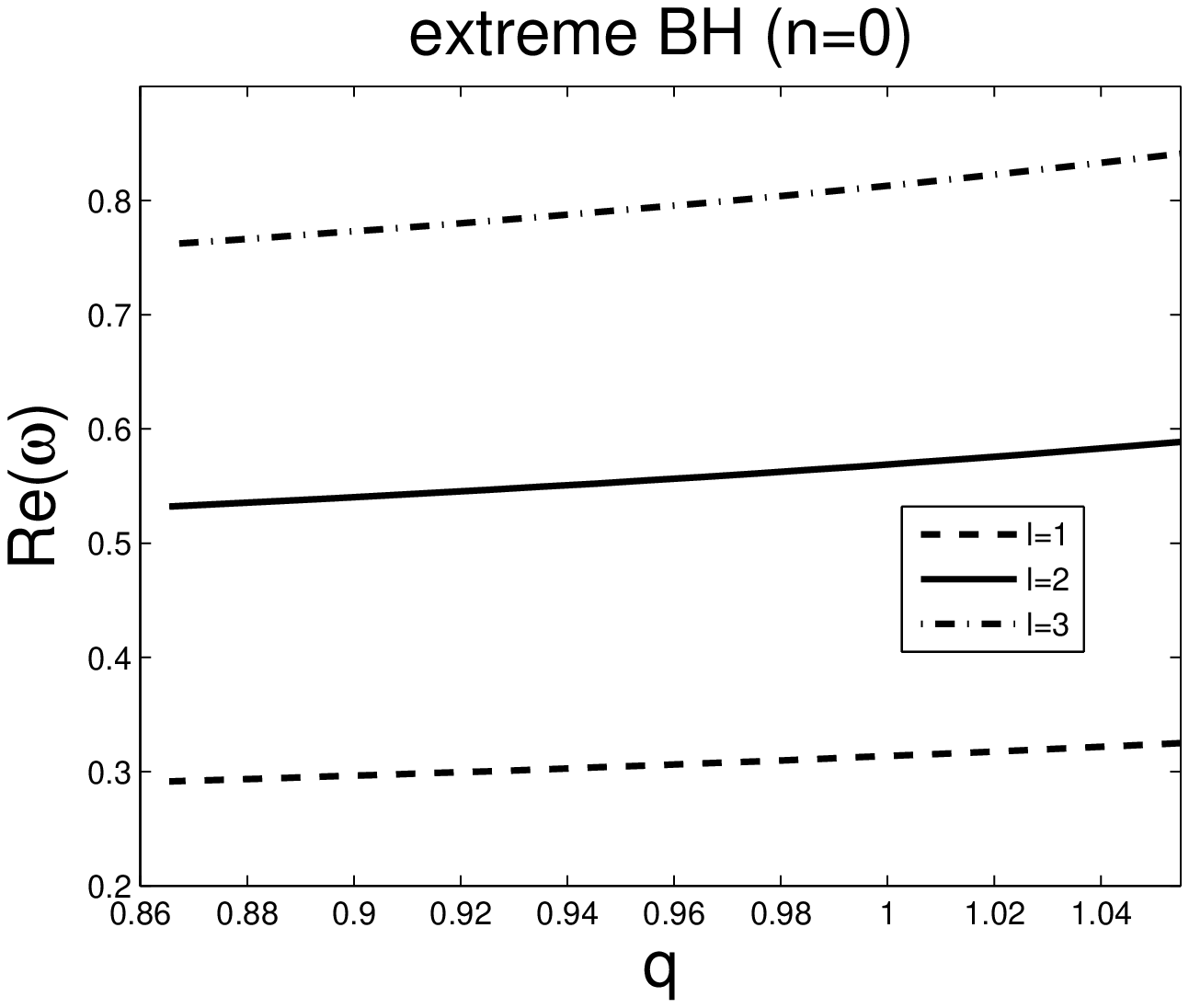}
\end{minipage}
} \subfigure{
\begin{minipage}[t]{0.4\textwidth}
\includegraphics[width=1\textwidth]{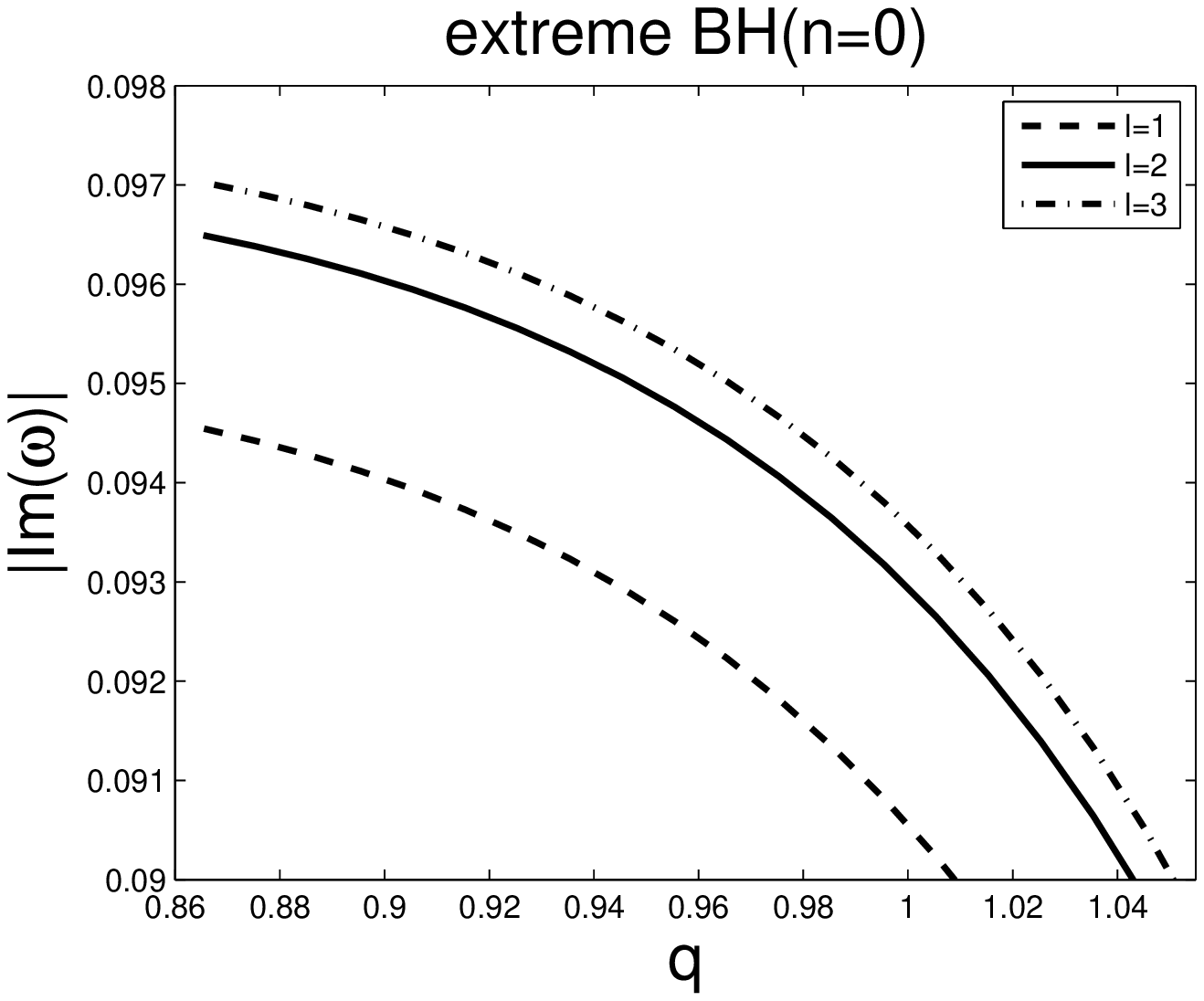} \\
\end{minipage}
}
\centering \subfigure{
\begin{minipage}[t]{0.4\textwidth}
\includegraphics[width=1\textwidth]{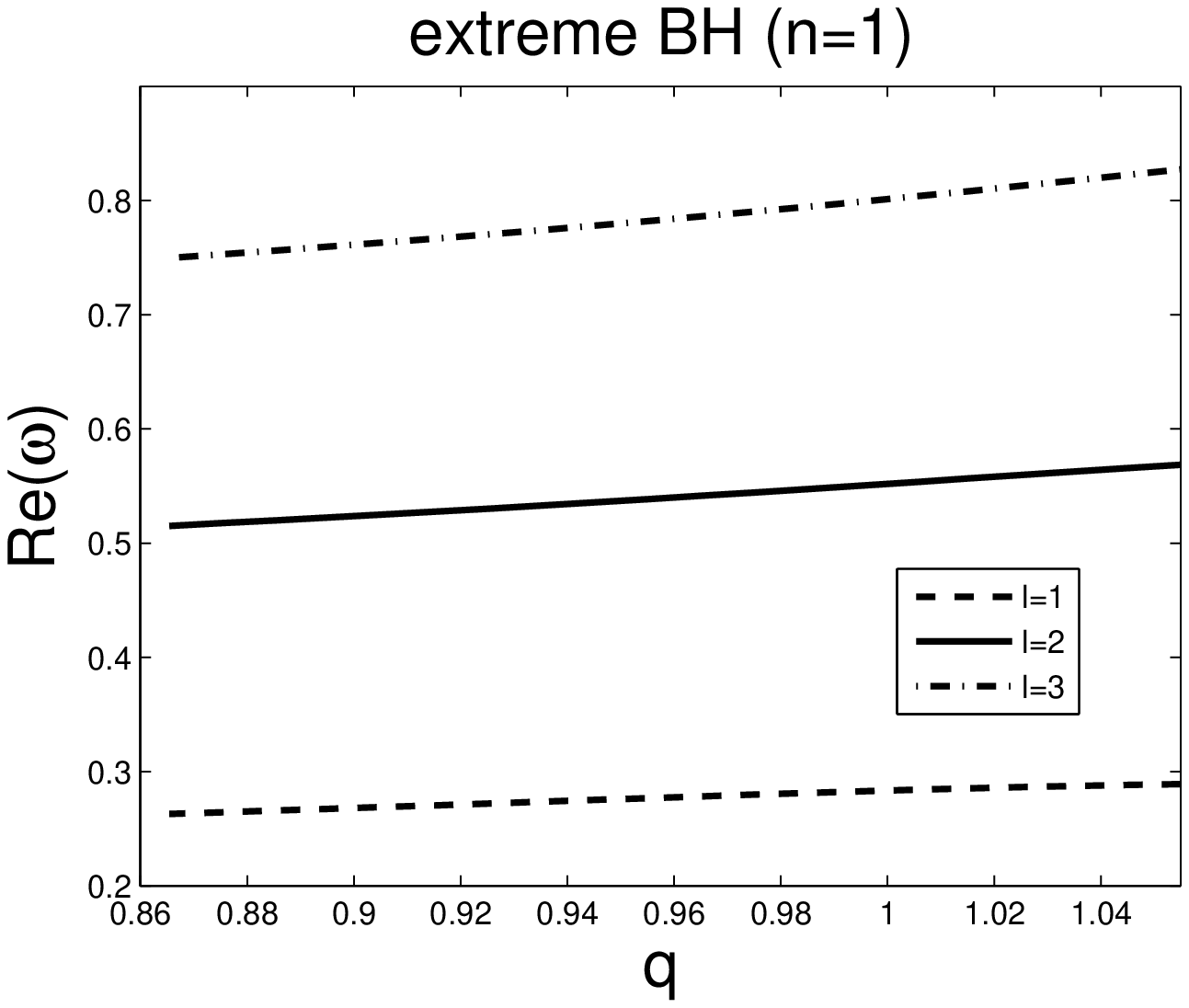}
\end{minipage}
} \subfigure{
\begin{minipage}[t]{0.4\textwidth}
\includegraphics[width=1\textwidth]{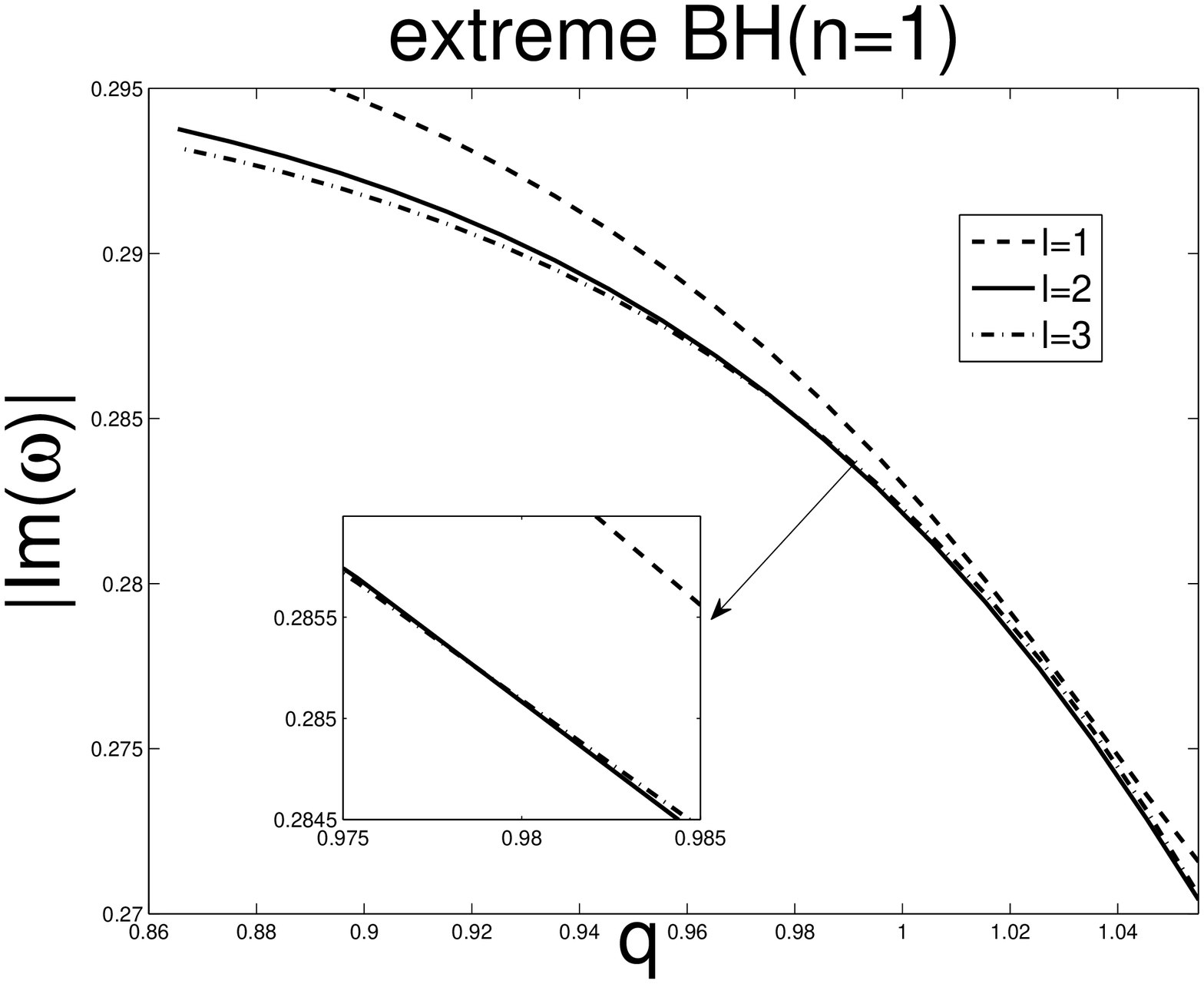} \\
\end{minipage}
}

\centering \subfigure{
\begin{minipage}[t]{0.4\textwidth}
\includegraphics[width=1\textwidth]{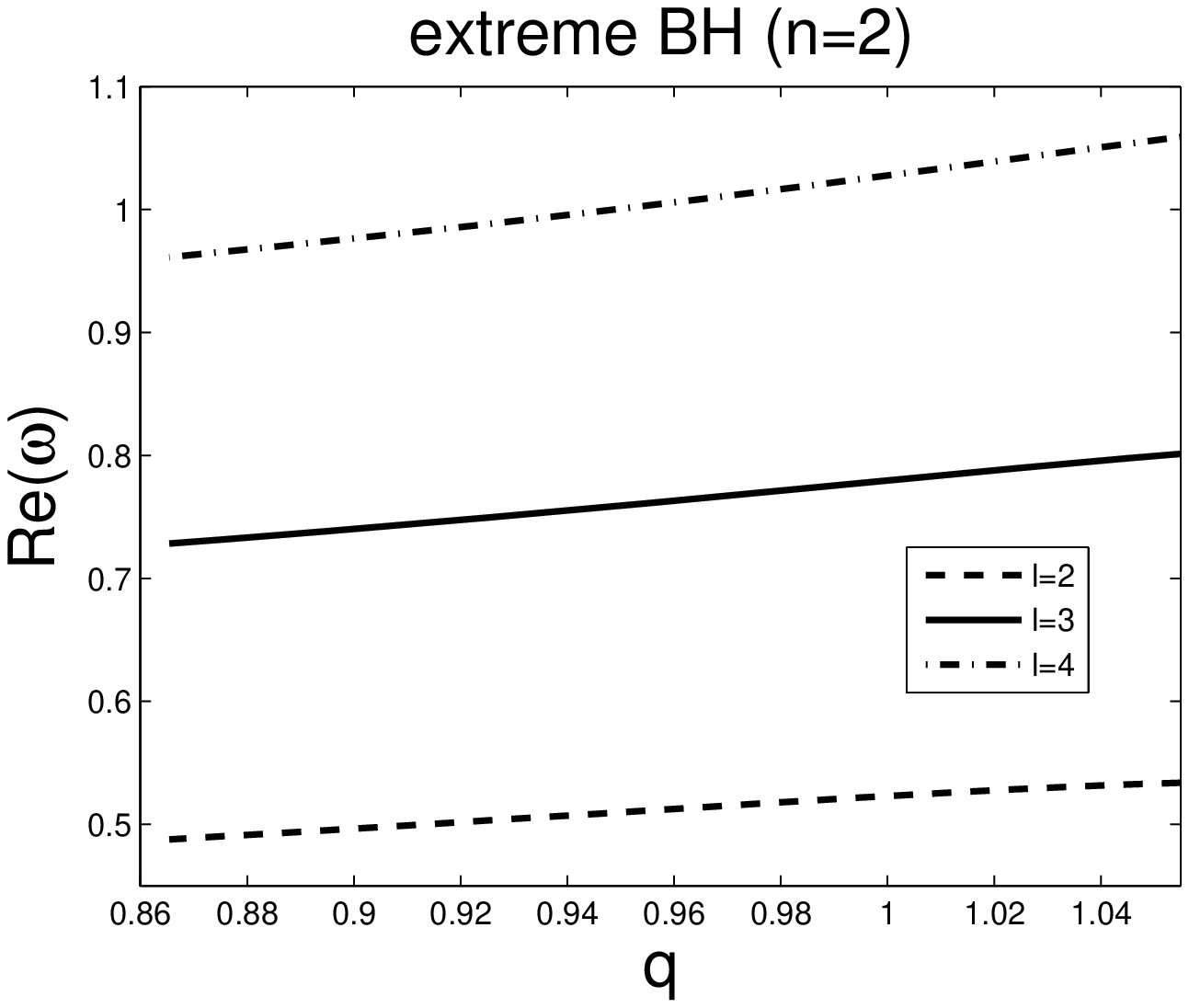}
\end{minipage}
} \subfigure{
\begin{minipage}[t]{0.4\textwidth}
\includegraphics[width=1\textwidth]{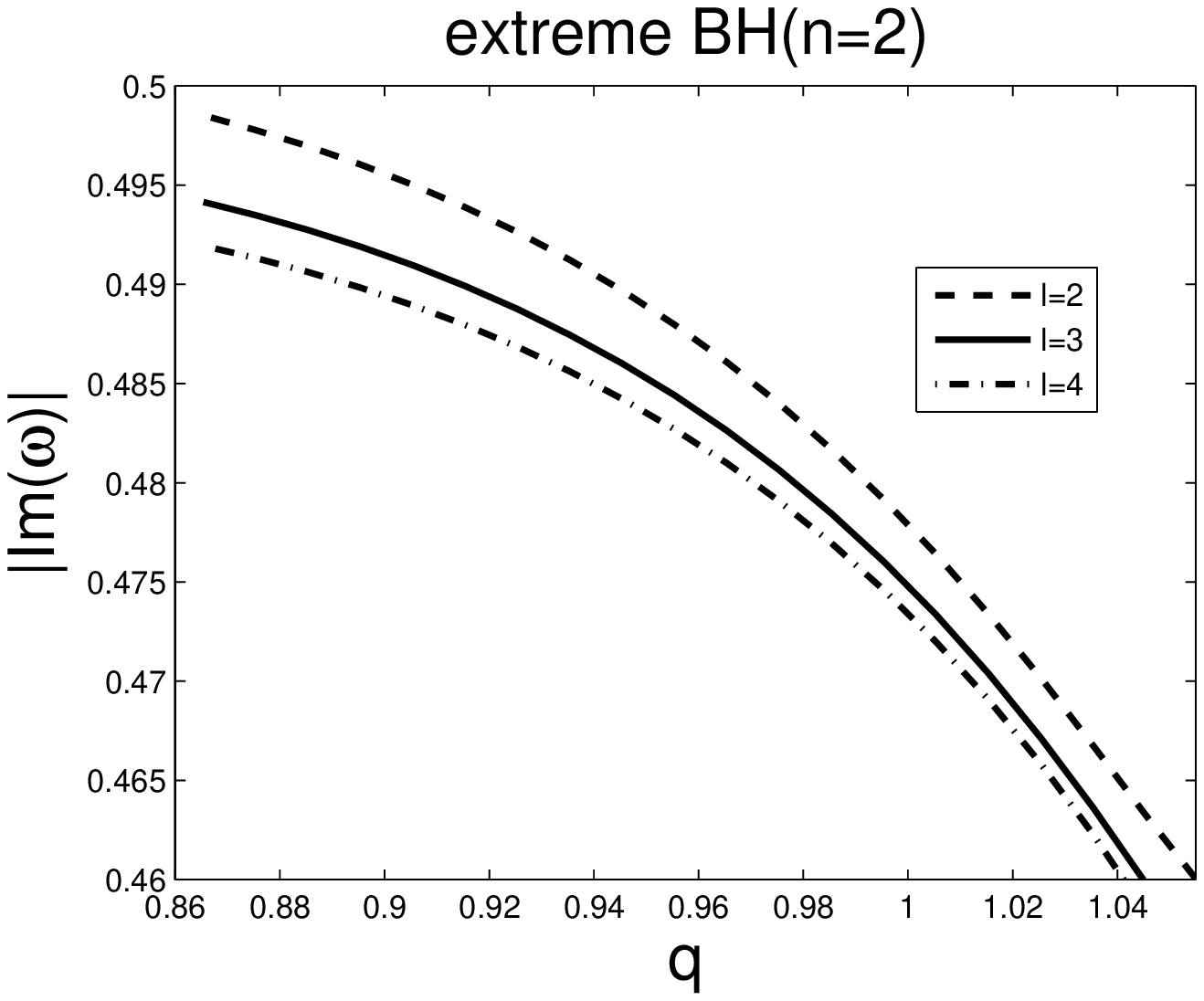} \\
\end{minipage}
}
\caption{The QNMs frequencies of the regular BH with strong charge}
\label{fig:QNMex(r)}
\end{figure}

FIG.\ref{fig:QNMex(r)} describes the QNMs frequencies of the regular BH around the extreme charge $q_{c}$. Like the behavior of weak charged cases, the real part of $\omega$ is enhanced by the increase of $q$ and $l$. But the imaginary part of $\omega$ behaves much differently from the weak charged conditions. It is found that the absolute value of $\rm{Im}(\omega)$ decreases with larger $q$ and lower overtone. The most important property is the effect of the angular momentum number, which increases the decay rate for the $n=0$ case, but reduces the decay speed when $n=2$. Especially in the $n=1$ case, $|\rm{Im}(\omega)|$ has the similar shape as the one with $n=2$ at lower charge. As the charge approaches $q_{c}$, the decay behavior is different from the cases of $n=0,2$. This may become the characteristic of such near-extreme BH's perturbation.

\begin{figure}
\centerline{\includegraphics[height=6cm]{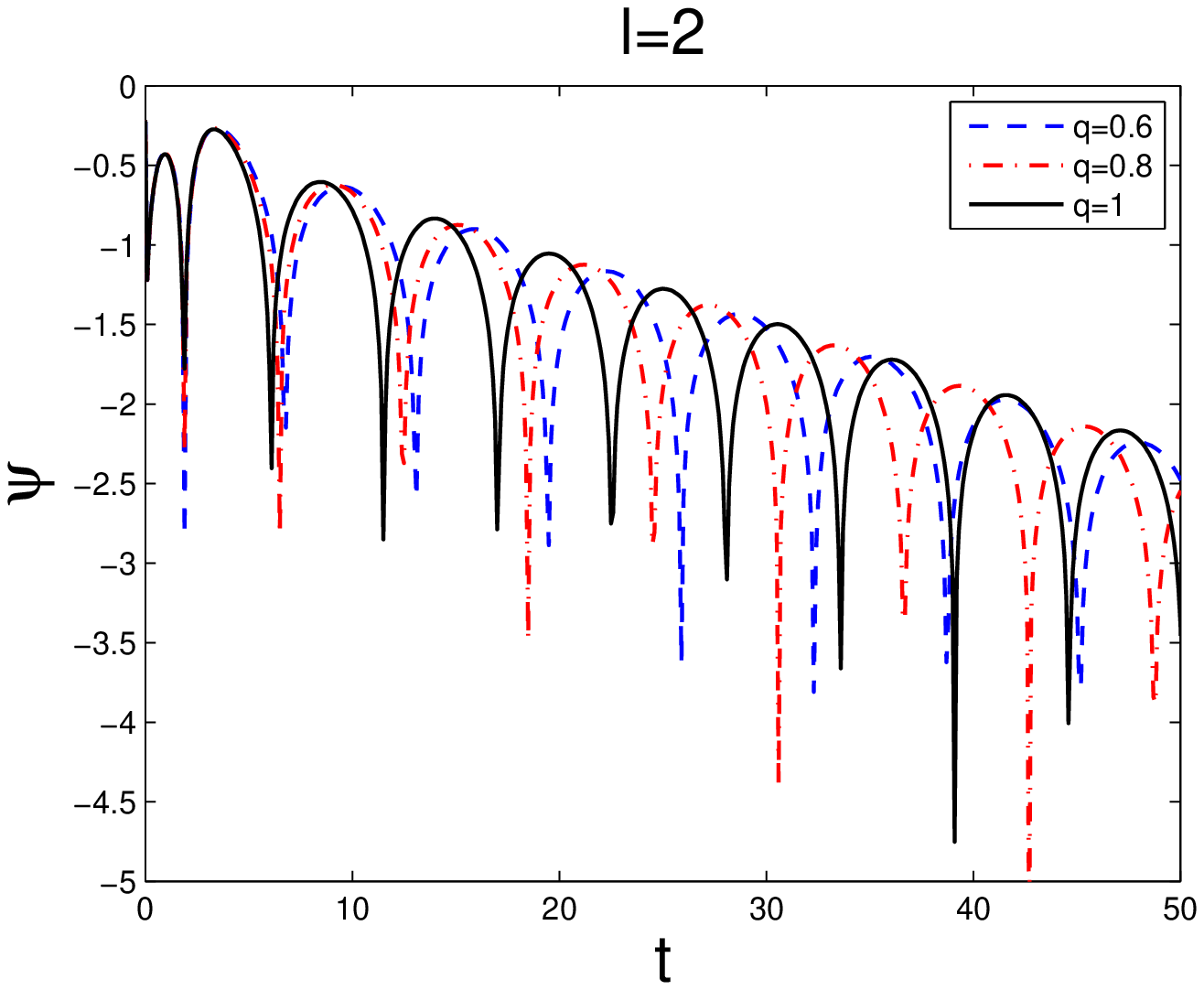}\includegraphics[height=6cm]{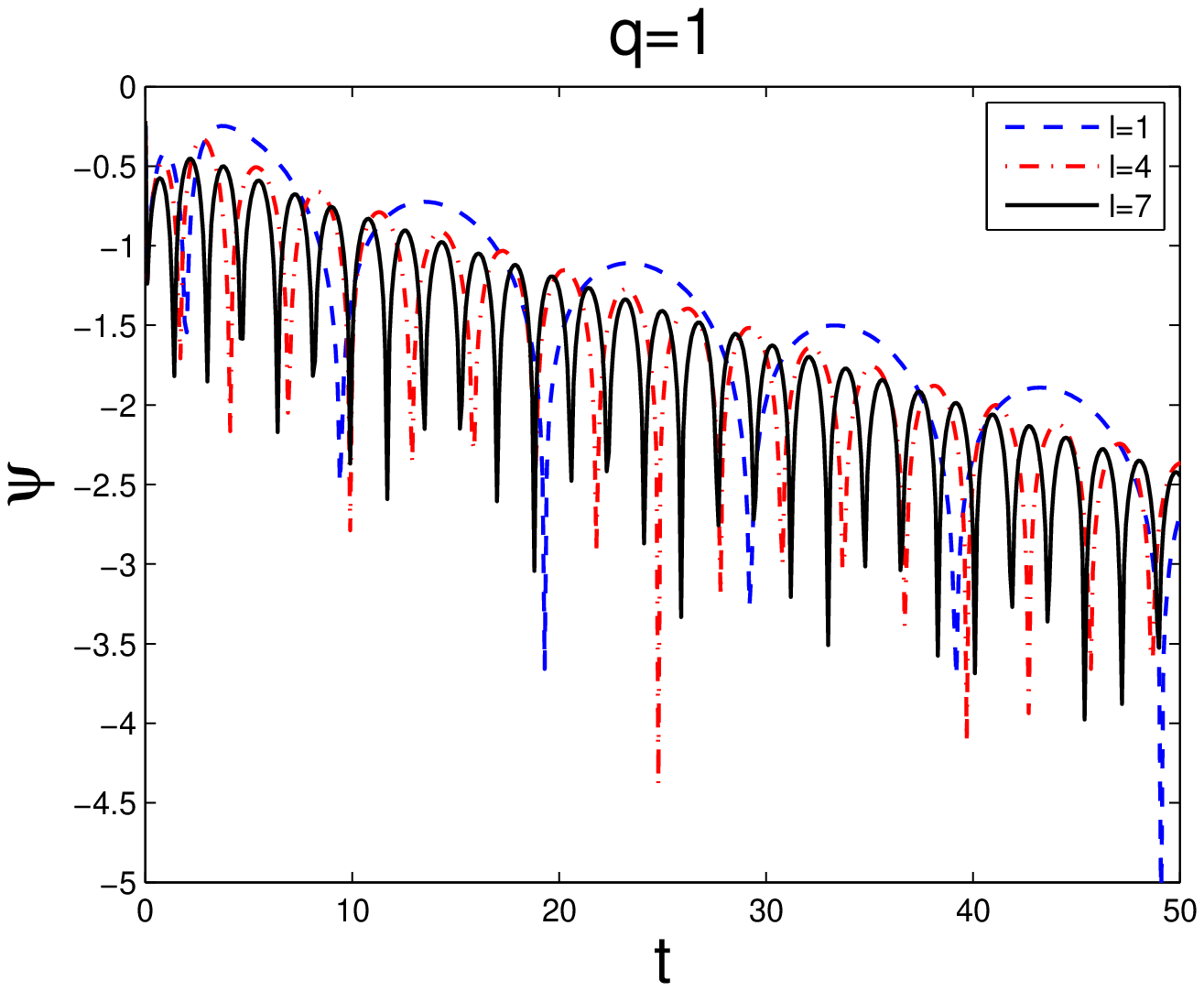}}
\caption{The dynamical evolution of NLED EM field in the background of the extreme regular black
hole spacetime. We set $M=1$. The
constants in the gaussian pulse $\nu_{c}=1$ and $\sigma=1$.}
\label{Fig:finitex}
\end{figure}

For the near-extreme cases, we also employ the finite difference method to determine the dynamical evolution of NLED EM perturbation. The same calculation procedures as mentioned in the previous section generate the results in FIG.\ref{Fig:finitex}. This figure shows that as $q$ increase the decay rate decreases while the oscillation frequency increases, which is also agreement with the WKB results.

\section{Hawking radiation of the magnetically charged regular black hole}\label{sec:Hawking}
As an important issue of black hole physics, Hawking radiation has been proven to support a new way to understand the thermodynamics of black holes.
Moreover recently an important relationship between QNMs and BH thermodynamics have been found\cite{ref1-6}-\cite{ref1-12}, which can be helpful to explain BH QNMs in terms of quantum levels. Such works successfully interpreted the quantization of a QNM through effective temperature\cite{ref1-9,ref1-10}, and found that as the electrons in Bohr model the QNMs frequency $\omega$ is also non-strictly continuous due to a quantum transition between two discrete energy levels. That is a brilliant interpretation of the Hawking radiation, but for our regular black hole it is difficult to find an analytic expression to calculate discrete QNMs frequency $\omega_{n}$ with $T_{\rm{eff}}$. So we apply a charged Hamilton-Jacobi equation, which is considered as an effective and simpler method for studying black hole's radiation, to study such regular black hole's Hawking radiation.

\begin{figure}
\centerline{\includegraphics[height=10cm]{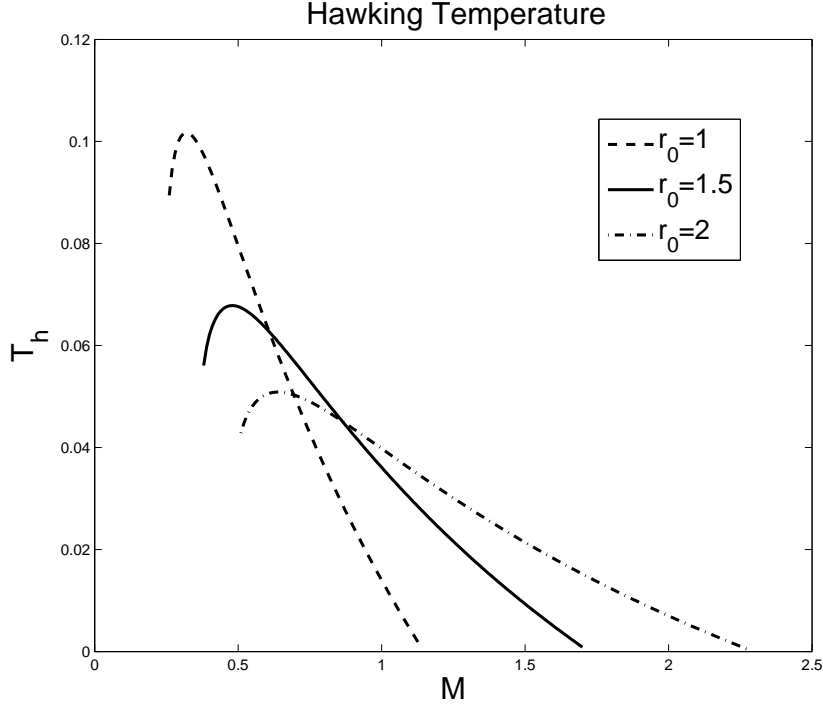}}
\caption{The Hawking temperature varies different mass of black hole. The dashing, solid, dash-dotted lines are $r_{0}=1,2,3$ respectively.}
\label{Fig:HawkingT}
\end{figure}

The charged Hamilton-Jacobi equation is given as~\cite{KLCPB}:
\begin{equation}
\label{HJeq1} g^{\mu\nu}\left(\frac{\partial S}{\partial
x^{\mu}}+eA_{\mu}\right)\left(\frac{\partial S}{\partial
x^{\nu}}+eA_{\nu}\right)+m^{2}=0,
\end{equation}
Where $S$ is the action and the $e$ is the charge of a Hawking radiation
particle, especially, in regular charged spacetime,
$A_\mu=\delta^3_\mu A_3=-q_m\delta^3_\mu \cos \theta$. Through
variable separation the action can be written as
 \begin{equation}
 S=-\omega t+R(r)+Y(\theta,\phi).
 \end{equation}
So in the regular black hole spacetime, Eq.(\ref{HJeq1}) can be rewritten as
\begin{equation}
\label{HJeq2}
-\frac{\omega^{2}}{f(r)}+f(r)\left(\frac{dR}{dr}\right)^{2}+m^{2}+\frac{\lambda}{r^{2}}=0,
\end{equation}
\begin{equation}
\label{HJeq3} \left(\frac{\partial
Y}{\partial\theta}\right)^{2}+\frac{1}{\sin^2\theta}\left(\frac{\partial
Y}{\partial\phi}-e q_m\cos\theta\right)^{2}=\lambda.
\end{equation}
According to Eq.(\ref{HJeq2}) and Eq.(\ref{HJeq3}), the radial function can be deduced from the partial equation
\begin{equation}
\frac{dR(r)}{dr}=\pm\frac{\sqrt{\omega^{2}-f(r)(m^{2}+\frac{\lambda}{r^{2}})}}{f(r)}.
\end{equation}
Since our concern is around the horizon $r_{0}$, we can expand the $f(r)$ as
\begin{equation}
f(r_{0})=f'(r_{0})(r-r_{0})+f''(r_{0})\frac{(r-r_{0})^{2}}{2}+....
\end{equation}
Then the radial function is
\begin{equation}
\label{HJeq4}
R_{\pm}(r)=\pm\int\frac{\sqrt{\omega^{2}-f'(r_{0})(r-r_{0})(m^{2}+\frac{\lambda}{r^{2}})}}{f'(r_{0})(r-r_{0})}dr.
\end{equation}
Applying residue theorem to Eq.(\ref{HJeq4}), the radial function $R(r)$ should be
\begin{equation}
R_{\pm}(r)=\pm\frac{i\pi\omega}{f'(r_{0})}.
\end{equation}
So the tunneling rate of Hawking radiation is~\cite{tun4,tun6}
\begin{equation}
\Gamma=\frac{\exp(2\rm{Im}R_{+})}{\exp(2\rm{Im}R_{-})}=\exp\left(-4\pi\frac{\omega}{f'(r_{0})}\right),
\end{equation}
and the Hawking temperature should be
\begin{equation}
  T_{h}=\frac{f'(r_{0})}{4\pi}=\frac{2M+(r_{0}-4M)\tanh^{-1}\left(1-\frac{r_{0}}{2M}\right)}{8M \pi r_{0}},
\end{equation}
where $r_{0}$ can be determined by
\begin{equation}\label{qm-T}
q^{2}_{m}=2Mr_{0}\tanh^{-1}\left(\frac{2M-r_{0}}{2M}\right).
\end{equation}
For the extreme BH, the Hawking temperature is zero. But in the weak charged cases, the Hawking temperature illuminates how the parameters, such as charge and mass of BH, impact on the thermodynamics of a black hole. FIG.\ref{Fig:HawkingT} shows the shapes of $T_{h}(M)$. In order to make the function $T_{h}(M,r_{0})$ meaningful, the domain of the mass shifts to the right with increasing $r_{0}$. For the cases with $r_{0}=1,1.5,2$, the Hawking temperature has a maximum value. Meanwhile, the figure tells us the relationship among $q_{m}, M$ and $T_{h}$.
\section{conclusions}\label{sec:conclusions}
\begin{figure}
\centering \subfigure{
\begin{minipage}[t]{0.3\textwidth}
\includegraphics[width=1\textwidth]{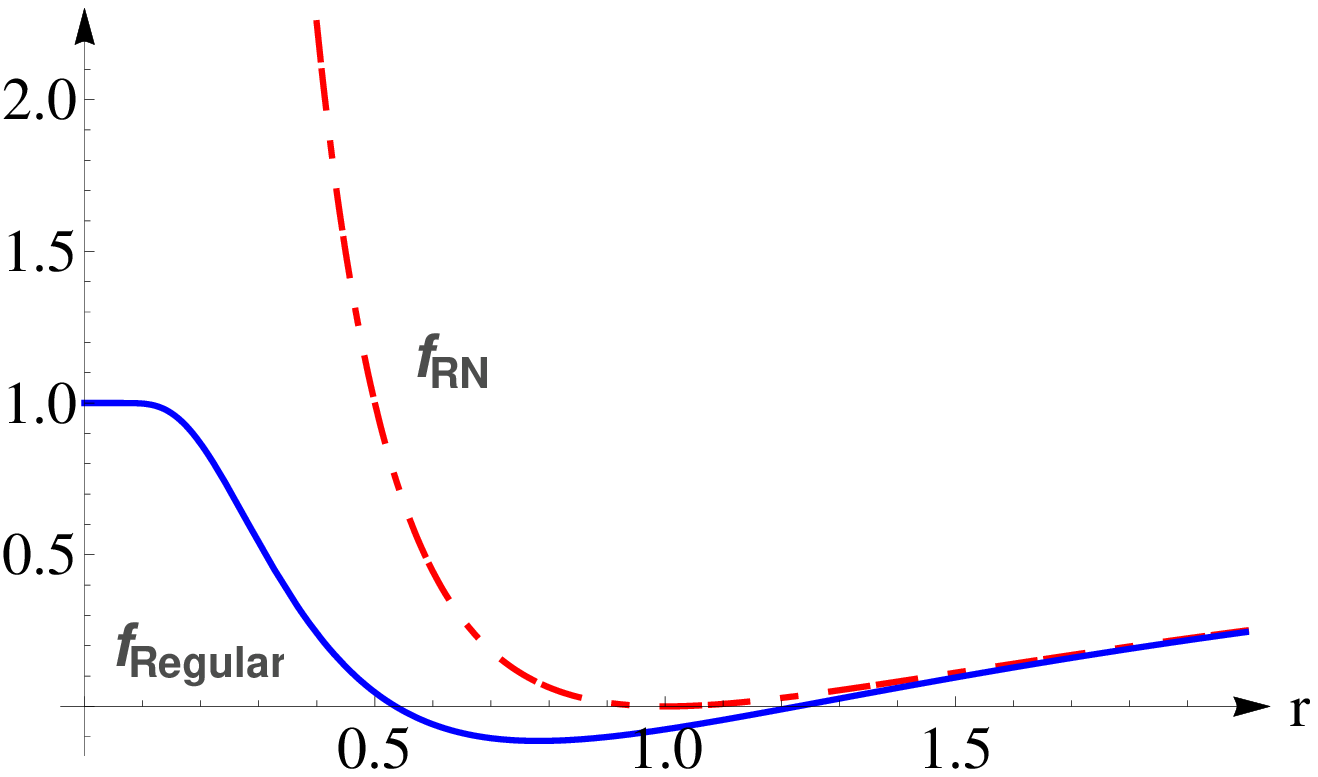}
\end{minipage}
} \subfigure{
\begin{minipage}[t]{0.3\textwidth}
\includegraphics[width=1\textwidth]{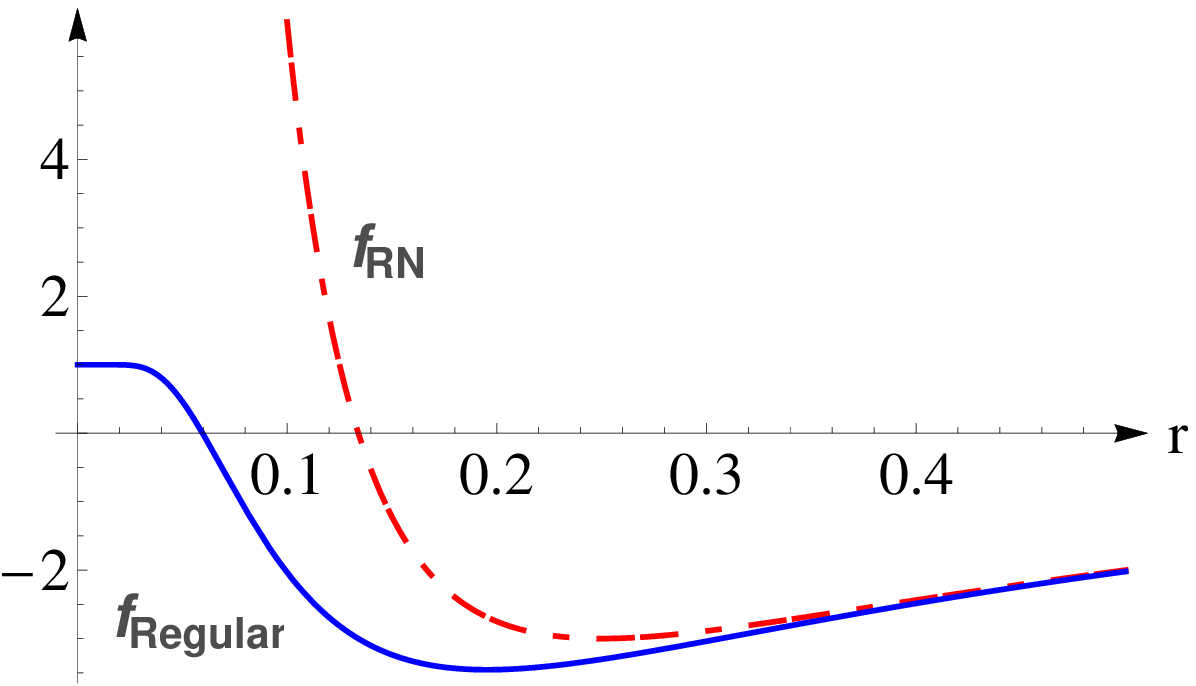}
\end{minipage}
} \subfigure{
\begin{minipage}[t]{0.3\textwidth}
\includegraphics[width=1\textwidth]{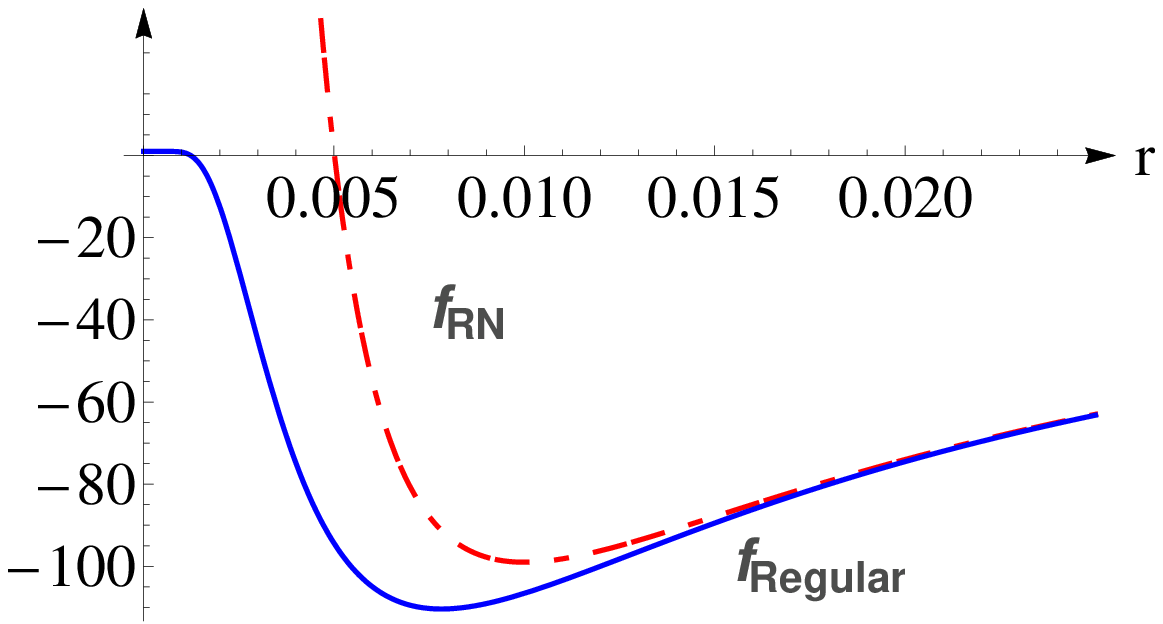}
\end{minipage}
} \caption{The figure illustrates $f(r)$ of
the regular metric and RN BH. From left to right $q=1,0.5,0.1$ respectively. Note for regular BH $q=q_{m}$ while
$q=q_{e}$ in RN condition}
\label{fig:f(r)}
\end{figure}

The regular BH discussed by us is one solution of nonlinear
electrodynamics coupling with Einstein theory. Choosing different Lagrangians
functions ${\cal L}(F)$, some other various regular solutions can be deduced~\cite{ourpaper}.
All of them can return to Schwarzschild black hole when $q=0$. For
non-zero $q$, they have no singularity---even at $r=0$, $f(r)$ is
finite. Furthermore, it can be expanded into
\begin{equation}
f(r)=1-\frac{2M}{r}+\frac{q^{2}}{r^{2}}-\frac{q^{6}}{12(M^{2}r^{4})}+{\cal
O}(q^{7}).
\end{equation}
In fact, all regular BHs can be expanded into a similar polynomial, so it can be noted that the regular metric asymptotically behaves as the RN BH. FIG.\ref{fig:f(r)} indicates the difference between the metric of regular BH and RN BH. It is apparent that the regular BHs with lower charge would be much more similar to RN space-time except in the area around the singular point $r=0$. In our work, we consider the electromagnetic perturbation as a test field, which is so weak that the background spacetime can be regarded as a static gravitational field. Namely, what we researched is probe limit case\cite{finaladd1}-\cite{finaladd4}. The results in above sections have no divergence term, so the probe limit condition could be satisfied. In appendix B, the general perturbation equations including gravitational and electromagnetic perturbation are derived (i.e. Eq.(\ref{perG}) and (\ref{perEM})).

For the weak charged conditions, the magnetic charge $q$ should be confined by $q<q_{c}$ to make sure that the regular spacetime has physical meaning. Considering the complicated expression of potential function, we try to seek more accurate QNMs frequency through higher order Taylor expansion. From $6^{th}$ WKB results FIG.\ref{fig:QNMReg(r)} shows the behavior of the perturbation field evolution in detail.

For the near-extreme cases, it is meaningful to reveal how the perturbation behaves when the magnetic charge $q$ approaches $q_{c}$. We find the distinguishing properties between near extreme case and weak charged conditions, which are mainly be reflected in the relationship between magnetic charge and decay rate. There is a curve intersection in the $n=1$ case, and the effect of angular momentum number $l$ on perturbation field decay appears opposite between $n=0$ and $n=2$.

But in each case, we can find the fundamental mode with lower angular momentum number plays a dominant role on perturbation field decay rate. This is also supported by the intuitive image from finite difference method.

From the discussion of Hawking radiation, we find another useful property of regular BH. Although we do not determine how the magnetic charge impacts on the temperature evidently, that can be deduced from Eq.(\ref{qm-T}).

As an important application of NLED theory, we choose this solution to understand the stability of initially global regular configurations. In one sense, it provides a new approach to improve general relativity. However,
we also notice that there are some deficiencies of this metric. For instance the Lagrangians ${\cal L}(F)$ is not a monotonic function in the whole range of invariant $F$, but rather having some junctions where singularities appear~\cite{KAB2}-\cite{KAB4}. That can be well revealed using the effective metric\cite{add3}. Furthermore some researchers put forward more generic conditions \cite{add1} where the inner horizons of such regular black holes would be instable due to a mass-inflation \cite{add2}. In addition to the specific regular solution discussed in this paper, there is another important non-singular solution for the gravitational collapse found in 2010 according to the relationship between NLED gravitational redshift and mass-radius ratio\cite{ref1-3}. Furthermore some non-singular cosmologies can also be found with NLED\cite{ref1-4,ref1-5}. Therefore we plan to study these issues in our future works.

\section*{\bf APPENDIX A}
\begin{figure}
\centering \subfigure{
\begin{minipage}[t]{0.4\textwidth}
\includegraphics[width=1\textwidth]{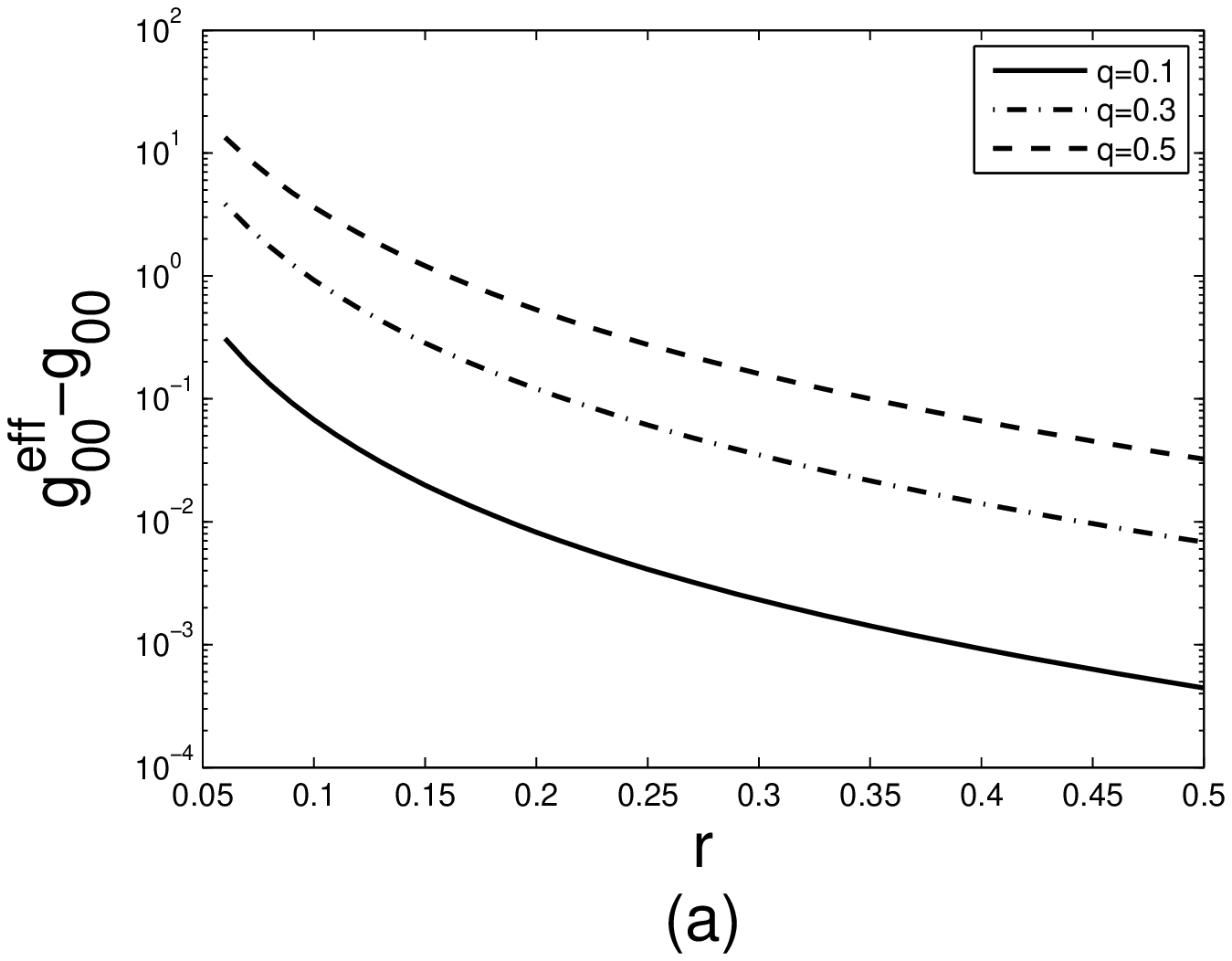}
\end{minipage}
} \subfigure{
\begin{minipage}[t]{0.4\textwidth}
\includegraphics[width=1\textwidth]{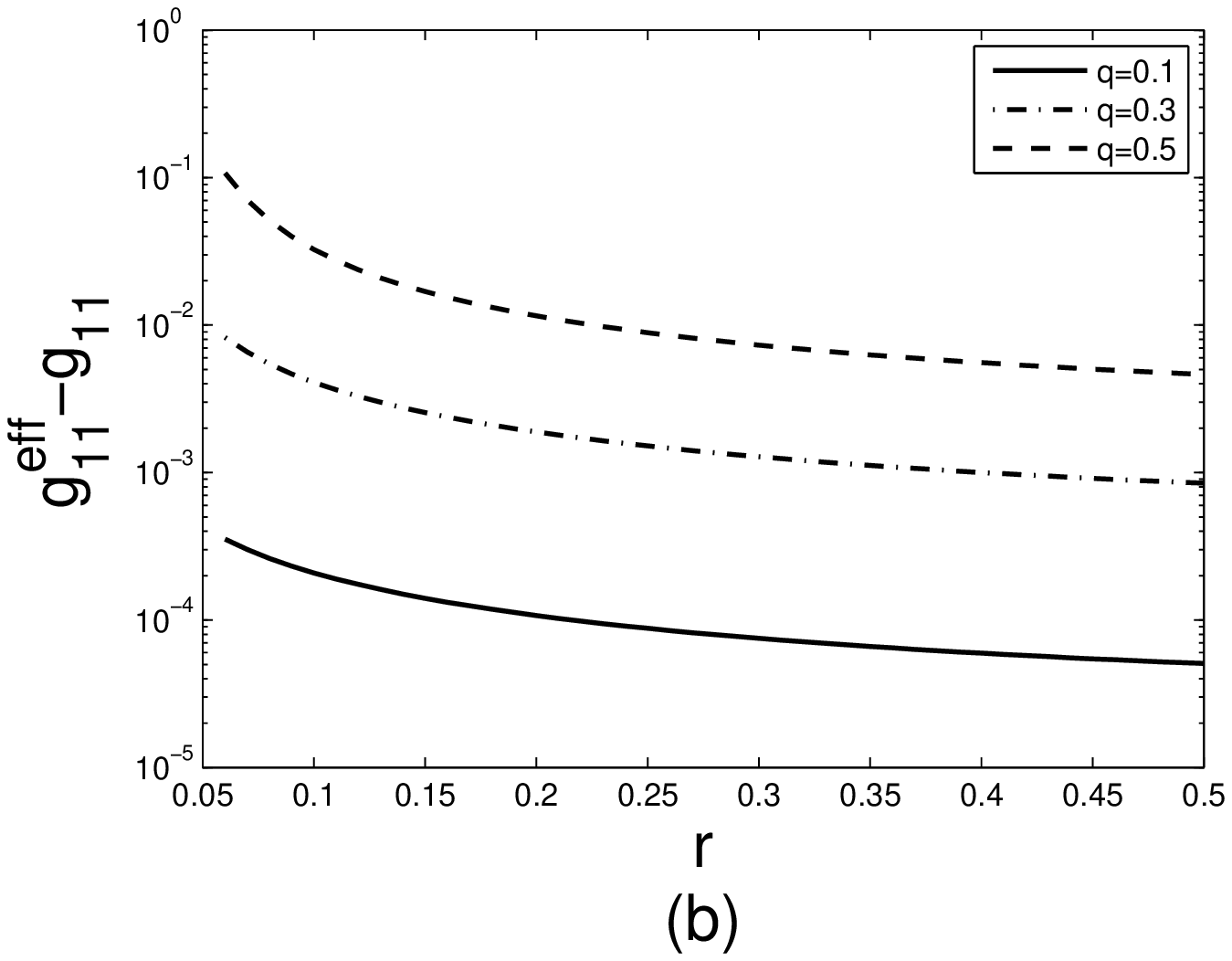} \\
\end{minipage}
}
\centering \subfigure{
\begin{minipage}[t]{0.4\textwidth}
\includegraphics[width=1\textwidth]{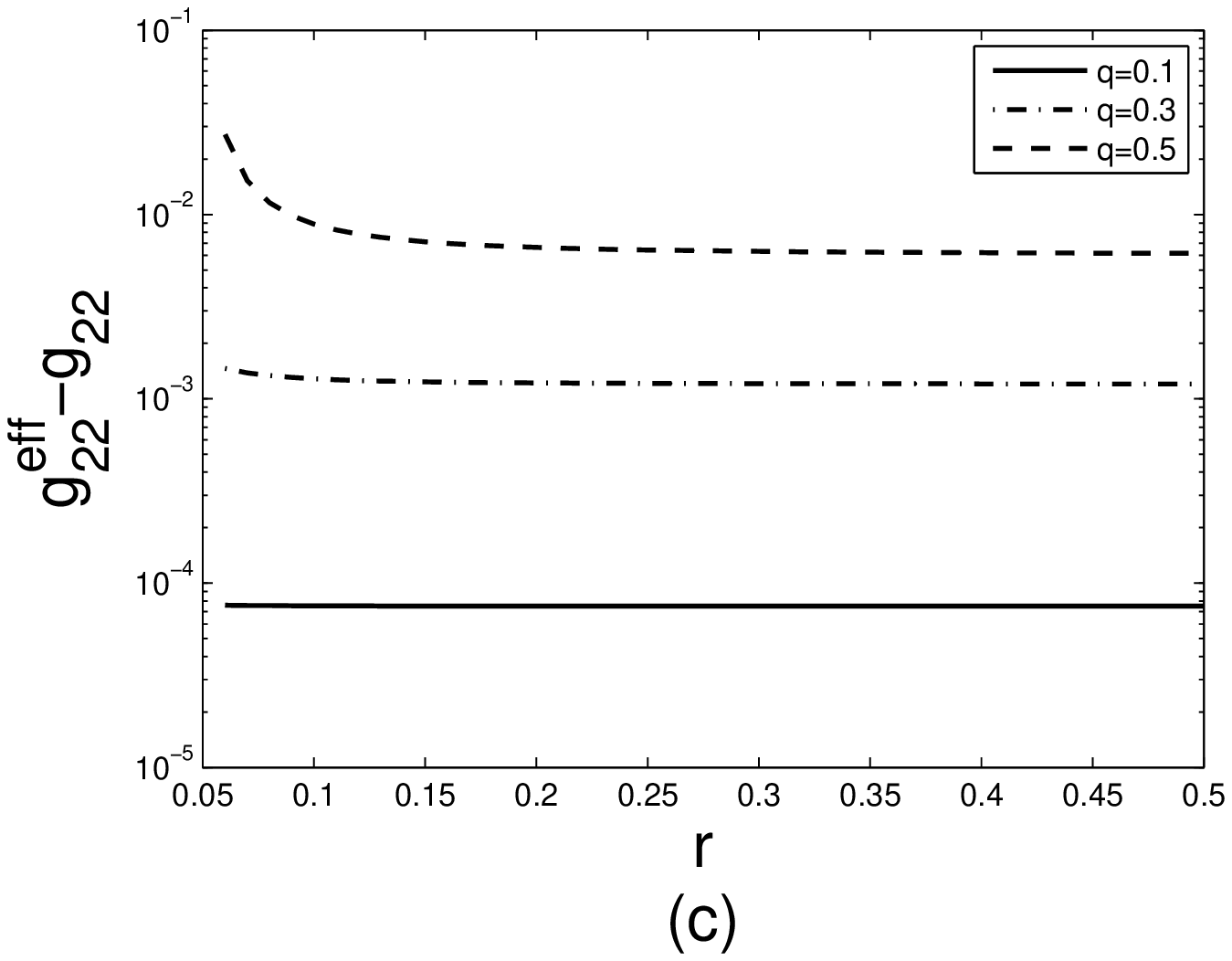}
\end{minipage}
} \subfigure{
\begin{minipage}[t]{0.4\textwidth}
\includegraphics[width=1\textwidth]{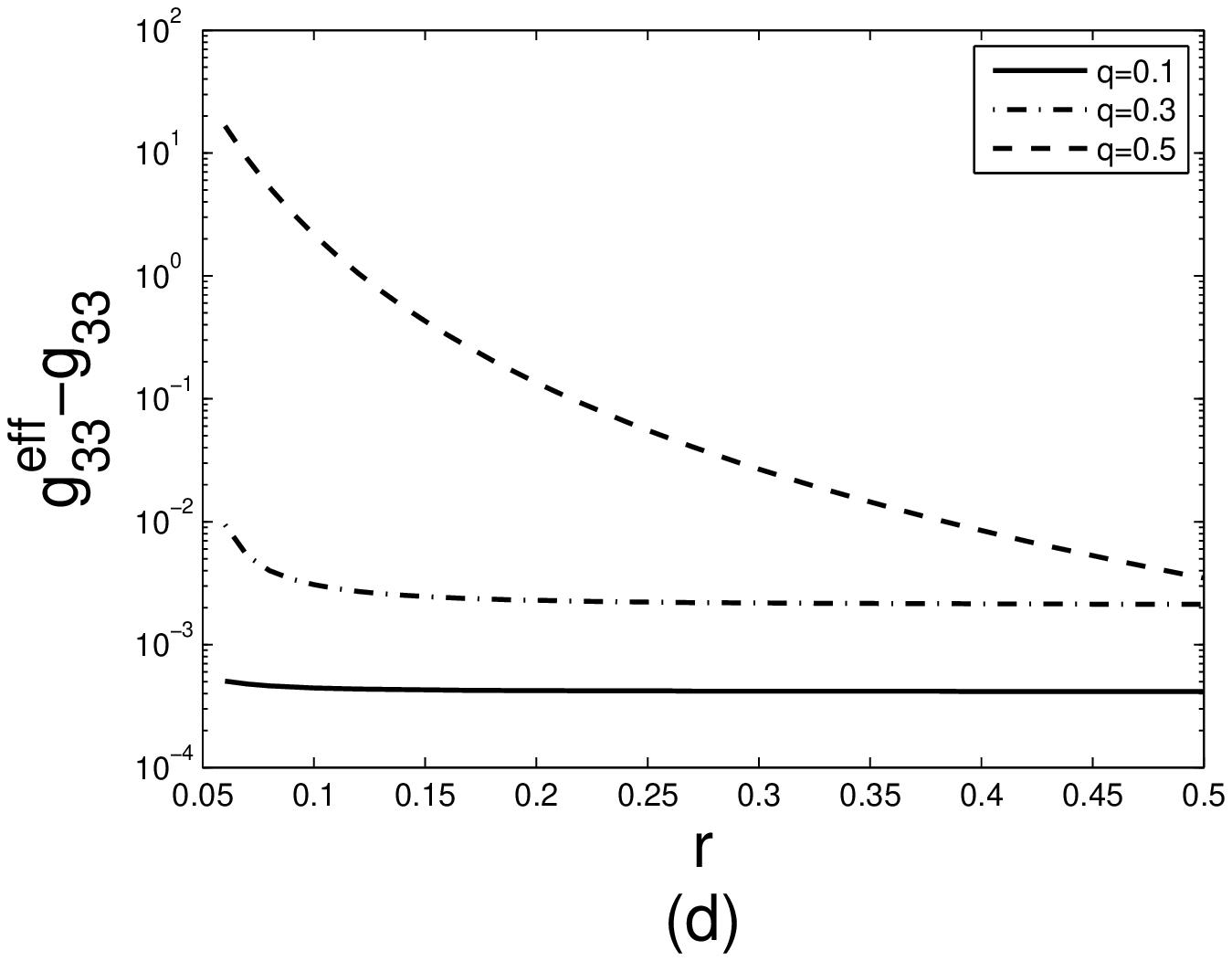} \\
\end{minipage}
}
\caption{The discrepancy between the effective metric and original metric.}
\label{fig:Geff}
\end{figure}

Taking our regular black hole into consideration, the effective metric due to NLED can be derived from the equation\cite{ref1-1}
\begin{equation}\label{geff}
g^{\mu\nu}_{\rm{eff}}={\cal L}_{\rm{F}}g^{\mu\nu}-4{\cal L}_{\rm{FF}}F^{\mu}_{\alpha}F^{\alpha\nu},
\end{equation}
where ${\cal L}_{\text{FF}}=d^{2}L/dF^{2}$. Then taking Eq.(\ref{eq:metric}), Eq.(\ref{L}) and (\ref{Fp}) into Eq.(\ref{geff}) yields
\ba &&
g^{00}_{\rm{eff}}=g^{00}+\frac{r\text{sech}^{2}(\frac{Q(r)}{2M})\left(Q(r)+2M\text{sinh}(\frac{Q(r)}{M})\right)\text{tanh}(\frac{Q(r)}{2M})}{4M(r-2M+2M\text{tanh}(\frac{Q(r)}{2M}))},
\ea
\ba &&
g^{11}_{\rm{eff}}=g^{11}-\frac{\text{sech}^{2}(\frac{Q(r)}{2M})\left(Q(r)+2M\text{sinh}(\frac{Q(r)}{M})\right)\text{tanh}(\frac{Q(r)}{2M})\left(r-2M+2M\text{tanh}(\frac{Q(r)}{2M})\right)}{4Mr},
\ea
\ba &&
g^{22}_{\rm{eff}}=g^{22}+\frac{\text{sech}^{4}(\frac{Q(r)}{2M})\left[-2Q^{2}(r)+Q^{2}(r)\text{cosh}(\frac{Q(r)}{M})-M\text{sinh}(\frac{Q(r)}{M})\left(7Q(r)+4M\text{sinh}(\frac{Q(r)}{M})\right)\right]}{16M^{2}r^{2}},
\ea
\ba &&
g^{33}_{\rm{eff}}=g^{33}-\frac{\text{csc}^{2}\theta\text{sech}^{4}(\frac{Q(r)}{2M})\left[\frac{q^{4}}{M^{2}r}\left(2-\text{cosh}(\frac{Q(r)}{M})\right)+2r\left(\text{cosh}(\frac{2Q(r)}{M})-1\right)+\frac{7q^{2}}{M}\text{sinh}(\frac{Q(r)}{M})\right]}{16r^{3}},
\ea
where $Q(r)=q^{2}/r$. Since the gravitational redshift is associated with the covariant form of effective metric, which can be derived from $g^{\mu\nu}_{\rm{eff}}g^{\rm{eff}}_{\nu\alpha}=\delta^{\mu}_{\alpha}$. Then we get
\ba &&
g^{\rm{eff}}_{00}=g_{00}+1+\frac{2M\left(\rm{tan}(\frac{Q(r)}{2M})-1\right)}{r}-\frac{4M\rm{cosh}(\frac{Q(r)}{2M})\left(2M\rm{tanh}(\frac{Q(r)}{2M})+r-2M\right)}{r\left(4M-Q(r)\rm{tanh}(\frac{Q(r)}{2})\right)},
\ea
\ba &&
g^{\rm{eff}}_{11}=g_{11}-1+\frac{4M}{r+r\exp(Q(r)/M)}+\frac{2Mr\left(1+\rm{cosh}(\frac{Q(r)}{M})\right)}{\left(4M-Q(r)\rm{tanh}(\frac{Q(r)}{2M})\right)\left(r-2M+2M\rm{tanh}(\frac{Q(r)}{2M})\right)},
\ea
\ba &&
g^{\rm{eff}}_{22}=g_{22}-\frac{1}{r^{2}}+\frac{8M^{2}q^{3/2}r^{3}\left(1+\rm{cosh}(\frac{Q(r)}{M})\right)}{2\frac{\sqrt{q}}{r}\left[8M^{2}qr^{\frac{7}{2}}+q^{5}\left(1-\frac{3}{1+\rm{cosh}(\frac{Q(r)}{M})}\right)\right]-14Mq^{\frac{7}{2}}\rm{tanh}(\frac{Q(r)}{2M})},
\ea
\ba &&
g^{\rm{eff}}_{33}=g_{33}-\frac{\rm{csc}^{2}\theta}{r^{2}}+\frac{16M^{2}q^{\frac{3}{2}}r^{4}\rm{cosh}^{2}(\frac{Q(r)}{2M})}{2\sqrt{q}\left(q^{5}+8M^{2}qr^{\frac{7}{2}}-\frac{3q^{5}}{1+\rm{cosh}(\frac{Q(r)}{M})}\right)-14Mq^{\frac{7}{2}}\rm{tanh}(\frac{Q(r)}{2M})}.
\ea
FIG.\ref{fig:Geff} illuminates the metric difference will be distinct around $r=0$ and tend to be eliminated with $r$ increase, meanwhile in the stronger charged cases the discrepancy becomes more obvious.
\section*{\bf APPENDIX B}
Considering the process of NLED EM perturbation to the regular spacetime, it should obey the following equations
\begin{equation}
  \delta G_{\mu\nu}=\delta T_{\mu\nu},
\end{equation}
\begin{equation}
\partial_{\nu} \delta(\sqrt{-g}{\cal L}_{F}F^{\mu\nu})=0,
\end{equation}
which are corresponding to the first order perturbation of Einstein field equation and NLED EM field equation (i.e., Eq.(\ref{3})) respectively, where $T_{\mu\nu}=-2{\cal L}_{F}g^{\alpha\beta}F_{\mu\alpha}F_{\nu\beta}+\frac{1}{2}g_{\mu\nu}{\cal L}$\cite{mainref}. Here the gravitational perturbations have been considered as $h_{\mu\nu}$, so we having
\begin{eqnarray}
 g_{\mu\nu} &=& \bar{g}_{\mu\nu}+h_{\mu\nu}.
\end{eqnarray}
Then
\ba &&
\delta G_{\mu\nu}=\nabla^{\beta}\nabla_{\mu}h_{\beta\nu}-\frac{1}{2}\nabla^{2}h_{\mu\nu}-\frac{1}{2}\nabla_{\mu}\nabla_{\nu}h-\frac{1}{2}\bar{g}_{\mu\nu}\nabla^{\alpha}\nabla^{\beta}h_{\alpha\beta} +\frac{1}{2}\bar{g}_{\mu\nu}\nabla^{2}h-\frac{\bar{R}}{2}h_{\mu\nu}-\frac{\bar{g}_{\mu\nu}}{2}\bar{R}_{\alpha\beta}h^{\alpha\beta}.\ea

According to Eq.(\ref{L}) and assuming $A_{\mu}=\bar{A}_{\mu}+\delta A_{\mu}$, we can get
\begin{equation}
\delta F=-2h^{\alpha\beta}\bar{F}_{\alpha\mu}\bar{F}^{\mu}_{\beta}+2\bar{F}^{\mu\nu}(\partial_{\mu}\delta A_{\nu}-\partial_{\nu}\delta A_{\mu}),
\end{equation}
\begin{eqnarray}
\delta {\cal L}=\bar{\cal L}_{F}\delta F,  ~~~~ \delta {\cal L}_{F}=\bar{\cal L}_{FF}\delta F.
\end{eqnarray}
Therefore
\ba &&
\delta T_{\mu\nu}=\left(4\bar{{\cal L}}_{FF}\bar{F}^{b}_{\mu}\bar{F}_{\nu b}\bar{F}_{\alpha c}\bar{F}^{c}_{\beta}-\bar{g}_{\mu\nu}\bar{{\cal L}}_{F}\bar{F}_{\alpha c}\bar{F}^{c}_{\beta}+\frac{1}{2}\bar{g}_{\mu\alpha}\bar{g}_{\nu\beta}\bar{{\cal L}}\right)h^{\alpha\beta}+2\left(\bar{g}_{\mu\nu}\bar{{\cal L}}_{F}\bar{F}^{\alpha\beta}-4\bar{F}^{\alpha\beta}\bar{{\cal L}}_{FF}\bar{F}^{b}_{\mu}\bar{F}_{\nu b}
\right. \nonumber\\ && \left.
+\bar{{\cal L}}_{F}\bar{F}^{\alpha}_{\nu}\delta^{\beta}_{\mu}-\bar{{\cal L}}_{F}\bar{F}^{\beta}_{\nu}\delta^{\alpha}_{\mu}+\bar{{\cal L}}_{F}\bar{F}^{\alpha}_{\mu}\delta^{\beta}_{\nu}-\bar{{\cal L}}_{F}\bar{F}_{\mu}^{\beta}\delta_{\nu}^{\alpha}\right)\partial_{\alpha}\delta A_{\beta}.
\ea
The general Einstein field equation with first order perturbation can be expressed in the form as
\ba\lb{perG} &&
\nabla^{\beta}\nabla_{\mu}h_{\beta\nu}-\frac{1}{2}\nabla^{2}h_{\mu\nu}-\frac{1}{2}\nabla_{\mu}\nabla_{\nu}h-\frac{1}{2}\bar{g}_{\mu\nu}\nabla^{\alpha}\nabla^{\beta}h_{\alpha\beta} +\frac{1}{2}\bar{g}_{\mu\nu}\nabla^{2}h-\frac{\bar{R}}{2}h_{\mu\nu}-\frac{\bar{g}_{\mu\nu}}{2}\bar{R}_{\alpha\beta}h^{\alpha\beta}=
 \nonumber\\ &&
\left(4\bar{{\cal L}}_{FF}\bar{F}^{~b}_{\mu}\bar{F}_{\nu b}\bar{F}_{\alpha c}\bar{F}^{~c}_{\beta}-\bar{g}_{\mu\nu}\bar{{\cal L}}_{F}\bar{F}_{\alpha c}\bar{F}^{~c}_{\beta}+\frac{1}{2}\bar{g}_{\mu\alpha}\bar{g}_{\nu\beta}\bar{{\cal L}}\right)h^{\alpha\beta}+2\left(\bar{g}_{\mu\nu}\bar{{\cal L}}_{F}\bar{F}^{\alpha\beta}-4\bar{F}^{\alpha\beta}\bar{{\cal L}}_{FF}\bar{F}^{~b}_{\mu}\bar{F}_{\nu b}
\right. \nonumber\\ && \left.
+\bar{{\cal L}}_{F}\bar{F}^{~\alpha}_{\nu}\delta^{\beta}_{\mu}-\bar{{\cal L}}_{F}\bar{F}^{~\beta}_{\nu}\delta^{\alpha}_{\mu}+\bar{{\cal L}}_{F}\bar{F}^{~\alpha}_{\mu}\delta^{\beta}_{\nu}-\bar{{\cal L}}_{F}\bar{F}_{\mu}^{~\beta}\delta_{\nu}^{\alpha}\right)\partial_{\alpha}\delta A_{\beta}.
\ea
Analogously the general NLED EM field equation with first order perturbation can be deduced as
\ba \lb{perEM} &&
\frac{\partial}{\partial x^{\nu}}\left\{\sqrt{-\bar{g}}\left[-\frac{1}{2}\bar{{\cal L}}_{F}\bar{F}^{\mu\nu}h-\left(\bar{{\cal L}}_{F}\bar{F}^{~\nu}_{\alpha}+\bar{{\cal L}}_{F}\bar{F}^{~\eta}_{\alpha}\delta^{\nu}_{\eta}+2\delta^{\beta}_{\eta}\bar{{\cal L}}_{FF}\bar{F}^{\eta\nu}\bar{F}_{\alpha m}\bar{F}^{~m}_{\beta}\right)h^{\mu\alpha}
\right.\right. \nonumber\\ && \left.\left.
+\left(4\bar{{\cal L}}_{FF}\bar{F}^{\mu\nu}\bar{F}^{\alpha\beta}+\bar{{\cal L}}_{F}\bar{g}^{\mu\alpha}\bar{g}^{\nu\beta}-\bar{{\cal L}}_{F}\bar{g}^{\mu\beta}\bar{g}^{\nu\alpha}\right)\partial_{\alpha}\delta A_{\beta}\right]
\right\}=0.
\ea

\section*{\bf APPENDIX C}
According to the transformation which proposed in \cite{NK3}, we discuss a source with electric and magnetic charges.The spacetime of the RN black hole with electric and magnetic charges can be expressed as\cite{NK1,NK2,NK3}
\ba \lb{metricNK} &&
ds^{2}=-\left(1-\frac{2Mr-(q^{2}_{e}+q^{2}_{m})}{r^{2}}\right)dt^{2}+\frac{r^{2}}{\Delta}dr^{2}+r^{2}d\theta^{2}+r^{2}\sin^{2}\theta d\phi^{2},
\ea
where $\Delta=r^{2}+(q^{2}_{e}+q^{2}_{m})-2Mr$. The electromagnetic vector potential and event horizon $r=r_{0}$ are\cite{NK0}
\bqn
A=-\frac{q_{e}}{r}dt-q_{m}\cos\theta d\phi;~~~~r_{0}=M+\sqrt{M^{2}-(q_{e}^{2}+q_{m}^{2})}.
\eqn
Since in this spacetime, electric and magnetic charges concentrate on the black hole, it is meaningful to combine the $q_{e}$ and $q_{m}$. Supposing that the densities of electric and magnetic charges satisfy $\rho_{e}/\rho_{m}=\cot\alpha$, where $\alpha$ is a real constant angle, then the Maxwell equation can be written as
\bqn
\nabla_{\nu}F^{\mu\nu}=4\pi\rho_{e}u^{\mu},~~~~~~~\nabla_{\nu}F^{+\mu\nu}=4\pi\rho_{m}u^{\mu},
\eqn
where $F^{+\mu\nu}$ is the dual tensor of $F^{\mu\nu}$, $u^{\mu}$ is the 4-velocity in curved spacetime. In order to construct an equivalent charge and the electromagnetic tensor, firstly we rewrite the electromagnetic tensor to be\cite{NK3,NK4}
\begin{equation}
\hat{F}^{\mu\nu}=F^{\mu\nu}\cos\alpha+F^{+\mu\nu}\sin\alpha.
\end{equation}
Yielding the equivalent Maxwell equation as
\bqn
\nabla_{\nu}\hat{F}^{\mu\nu}=4\pi(\rho_{e}\cos\alpha+\rho_{m}\sin\alpha)u^{\mu},~~~~\nabla_{\nu}\hat{F}^{\mu\nu}=4\pi(-\rho_{e}\sin\alpha+\rho_{m}\cos\alpha)u^{\mu}.
\eqn
Let $\alpha$ satisfy
\bqn
\rho_{e}\cos\alpha+\rho_{m}\sin\alpha=\rho_{h},~~~~\rho_{m}\cos\alpha-\rho_{e}\sin\alpha=0,
\eqn
so the equivalent charge density is $\rho_{h}=\sqrt{\rho^{2}_{e}+\rho^{2}_{m}}$, and the equivalent charge can be $q_{h}^{2}=q_{e}^{2}+q_{m}^{2}$. Then the equivalent electromagnetic vector potential should be
\begin{equation}
\hat{A}=\frac{q_{h}}{r}dt.
\end{equation}
Therefore the line element of the RN black hole with electric and magnetic charges can be simplified as
\begin{equation}
ds^{2}=-(1-\frac{2M}{r}+\frac{q^{2}_{h}}{r^{2}})dt^{2}+(1-\frac{2M}{r}+\frac{q^{2}_{h}}{r^{2}})^{-1}dr^{2}+r^{2}(d\theta^{2}+\sin^{2}\theta d\phi^{2}),
\end{equation}
which has the same expression as ordinary RN solution just letting $q_{h}=q_{e}$.
\section*{\bf Acknowledgements}
  We would like to express my gratitude to Professor Matthew Benacquista for his great help, and be grateful to the referees for their suggestions which are important for our paper's improvement. This work was supported by FAPESP No. 2012/08934-0, National Natural Science Foundation of China No. 11205254, No. 11178018 and No.
11375279, and the Natural Science Foundation Project of CQ CSTC
2011BB0052, and the Fundamental Research Funds for the Central Universities CQDXWL-2013-010 and CDJRC10300003.

\end{document}